\makeatletter
\declare@file@substitution{revtex4-1.cls}{revtex4-2.cls}
\makeatother

\documentclass[twocolumn]{aastex631}

\usepackage{newtxtext,newtxmath}
\usepackage[T1]{fontenc}
\usepackage{ae,aecompl}

\usepackage{mathtools} 
\usepackage{amsmath} 
\usepackage{bm} 
\usepackage{amsfonts} 
\usepackage{graphicx} 
\usepackage{bigstrut} 
\usepackage{tabularx} 
\usepackage{cancel} 
\usepackage{xcolor}
\usepackage{soul}
\usepackage{upgreek}
\usepackage{slashed} 
\usepackage{epstopdf} 
\newcommand\Lagr{\mathcal{L}} 

\newcommand\mue{\mu_{\rm e}} 
\newcommand\nmax{n_{\text{max}}} 
\usepackage[subnum]{cases} 
  {\color{red}}%
  {}

\shorttitle{Electron MHD with Landau-quantized electrons}
\shortauthors{P.~B. Rau and I. Wasserman}

\begin{document}


\title{Numerical simulation of electron magnetohydrodynamics with Landau-quantized electrons in magnetar crusts}

\correspondingauthor{Peter B. Rau}
\email{peter.rau@columbia.edu}

\author[0000-0001-5220-9277]{Peter B. Rau}
\affiliation{Institute for Nuclear Theory, University of Washington, Seattle, WA 98195-1550, USA}
\affiliation{Columbia Astrophysics Laboratory, Columbia University, New York, NY 10027, USA}

\author{Ira Wasserman}
\affiliation{Cornell Center for Astrophysics and Planetary Science, Cornell University, Ithaca, NY 14853, USA}

\begin{abstract}

In magnetar crusts, magnetic fields are sufficiently strong to confine electrons into a small to moderate number of quantized Landau levels. This can have a dramatic effect on the crust's thermodynamic properties, generating field-dependent de Haas--van Alphen oscillations. We previously argued that the large-amplitude oscillations of the magnetic susceptibility could enhance the Ohmic dissipation of the magnetic field by continuously generating small-scale, rapidly dissipating field features. This could be important to magnetar field evolution and contribute to their observed higher temperatures. To study this, we performed quasi-3D numerical simulations of electron MHD in a representative volume of neutron star crust matter, for the first time including the magnetization and magnetic susceptibility resulting from Landau quantization. We find that the potential enhancement in the Ohmic dissipation rate due to this effect can be a factor $\sim 3$ for temperatures of the order of $10^8$ K and $\sim 4.5$ for temperatures of the order of $5\times10^7$ K, depending on the magnetic field configuration. The nonlinear Hall term is crucial to this amplification: without it the magnetic field decay is only enhanced by a factor $\lesssim 2$ even at $5\times10^7$ K. These effects generate a high wavenumber plateau in the magnetic energy spectrum associated with the small-scale de Haas--van Alphen oscillations. Our results suggest that this mechanism could help explain the magnetar heating problem, though due to the effect's temperature-dependence, full magneto-thermal evolution simulations in a realistic stellar model are needed to judge whether it is viable explanation.

\end{abstract}


\section{Introduction}

The astrophysical implications of the quantization of the momentum of electron motion perpendicular to an applied magnetic field, or Landau quantization, have been studied for decades ~\citep{Canuto1971,Hernquist1984}. The strong magnetic fields of neutron stars in particular have become a focus for research in this area (see~\citet{Harding2006} for a review). Restricting the discussion to the interior of neutron stars, Landau quantization can significantly modify electron transport properties (e.g.,~\citet{Kaminker1981,Yakovlev1984,Hernquist1984,Potekhin1999a}), which become strongly anisotropic and undergo Shubnikov--de Haas (SdH) oscillations. This in turn changes neutron star thermal evolution (e.g.,~\citet{Hernquist1985,Potekhin2018}). Quantization additionally modifies neutrino emissivity and shifts the threshold density for the direct Urca reaction in the core~\citep{Leinson1998,Baiko1999,Maruyama2022}. The equation of state of the neutron star core, where other fermion species are also quantized in the presence of strong fields, is not strongly affected at physically relevant field strengths $B\lesssim 10^{18}$ G~\citep{Broderick2000}, though strong fields can shift the muon threshold density~\citep{Suh2001}. In the less-dense crust, these effects are more important, and can change the crustal nuclear composition~\citep{Chamel2012,Chamel2020} and move the neutron drip-line~\citep{Chamel2015}. Landau quantization also induces de Haas--van Alphen (dHvA) oscillations of the differential magnetic susceptibility of matter, which leads to magnetic domain formation within neutron stars~\citep{Blandford1982,Suh2010,Rau2021,Rau2023}.

In the crust, where Landau quantization effects are most pronounced, the magnetic field evolution is usually assumed to be governed by electron magnetohydrodynamics (EMHD), in which the electrons move with respect to an essentially static nuclear lattice, and the Hall term $\propto\bm{J}\times\bm{B}$ is dominant within Ohm's law. EMHD and the related Hall MHD are well-studied topics in neutron star crusts~\citep{Cumming2004,Pons2007,Pons2009,Vigano2012,Vigano2013,Gourgouliatos2013,Gourgouliatos2014,
Gourgouliatos2015,Lander2019,Brandenburg2020,Gourgouliatos2022,Vigano2021,Dehman2023}, including for magnetar-strength fields $\sim10^{14}$--$10^{16}$ G. However, the assumptions underlying EMHD possibly break down for very strong magnetic fields: if the Lorentz force exceeds the maximum elastic stress the crust can support, the crust can undergo plastic failure and the nuclear lattice can no longer be assumed stationary. A recent study~\citep{Gourgouliatos2021} has shown that plastic failure of the crust does not appear to completely suppress the Hall effect, and so a description of a magnetar crust's magnetic evolution based on EMHD should still be valid.

In a previous paper (\citet{Rau2023}, henceforth RW23), we examined MHD stability in a magnetar crust under the assumption that EMHD (there termed Hall MHD) is still applicable. Using linear mode analysis, we demonstrated the existence of a strong-field Hall MHD instability in a physically relevant region of magnetic field--density--temperature phase space. However, since this instability disappears upon magnetic domain formation, it is of limited astrophysical interest. We also found that Ohmic dissipation could be enhanced due to the oscillations of the differential magnetic susceptibility caused by Landau quantization. Unlike the oscillations of the magnetization, which are a small fraction $\lesssim1$\% of the magnetic field $\bm{B}$, the oscillations of $4\pi$ times the differential magnetic susceptibility can be of order unity in amplitude, comparable to the value of the susceptibility with zero magnetization. This effect is expected to persist for a wide range of temperatures, densities and field strengths, but a numerical study is needed to assess its overall importance to magneto-thermal evolution.

The effect of Landau quantization on magnetar thermal evolution has been considered previously in the context of modified thermal conductivity~\citep{Hernquist1985,Potekhin2018}, but these calculations have assumed a fixed magnetic field configuration. The joint magneto-thermal evolution of the crusts of magnetars with Landau quantization effects in both conductivities and thermodynamic quantities (e.g., magnetization and differential magnetic susceptibility) has not been considered before, for obvious reasons: the highly spatially inhomogeneous coefficients are difficult for numerical simulations to handle. There is good motivation to include these effects; however, magnetars are systematically observed to be more luminous than other neutron stars with surface luminosities $\mathcal{L}_s\sim 10^{35}$ erg/s, an order of magnitude or more greater than other neutron stars of the same age, and dissipation of their strong fields into heat is a likely explanation. Using magneto-thermal simulations,~\citet{Vigano2013} argued that the diversity of magnetars can be explained by a combination of (1) Ohmic dissipation of strong magnetic fields aided by the Hall effect, which generates small-scale field structure and hence promotes dissipation; (2) light element envelopes formed of accreted matter, which enhances thermal conductivity and supports a hotter surface. However, they included no Landau quantization effects in their calculation.

~\citet{Beloborodov2016} examined four possible mechanisms to explain magnetar heating: ambipolar diffusion heating of the core, mechanical dissipation of the crust stressed beyond the elastic limit, Ohmic dissipation of the crustal magnetic field and surface bombardment by accelerated charged particles. They found that the second and fourth mechanisms were insufficient to explain the observed luminosities, and that a magnetar heated by ambipolar diffusion in it core would not have a lifetime consistent with observations.~\citet{Tsuruta2023} recently re-examined the ambipolar diffusion heating scenario, finding that it could be consistent with observation in the case of light element contamination of the crust. For the crustal Ohmic dissipation scenario to be able to power the observed surface luminosities without a light element envelope,~\citet{Beloborodov2016} argued that fields changing by $\gtrsim 10^{16}$ G over length scales $\lesssim 10^2$ cm would be required. Notably the authors did not consider Landau quantization effects, and dHvA oscillations of magnetization and its derivatives could continuously generate rapidly dissipating small length-scale features in the magnetic field, enhancing its Ohmic dissipation and hence crustal heating.

In this paper, we perform quasi-3D numerical simulations of the magnetic induction equation for EMHD with Ohmic dissipation in a reference volume of neutron star crust matter. We include the effects of Landau quantization of electrons in the magnetization and its partial derivatives, neglecting for now the SdH oscillations of the electrical conductivity. To perform the simulations, we use the spectral method solver \texttt{Dedalus}, which has recently been used for studying the resistive instability of a particular choice of stellar magnetic field~\citep{Kaufman2022} and for simulating ambipolar diffusion in neutron star cores~\citep{Igoshev2023}. To make the numerics tractable, we use certain simplifying approximations that leave the key novel physics intact and allow us to assess its importance to magnetic field evolution. The most critical of these approximations is that EMHD is applicable throughout the simulation, which is not necessarily true if the field is strong enough to cause the crust to undergo plastic failure. The goal of this work is to demonstrate the effects of dHvA oscillations of magnetization and its partial derivatives on magnetic field evolution in a physically simplified setting. We determined the extent to which this effect increases Ohmic dissipation of the field and the generation of small-scale magnetic field structure, which further promotes Ohmic dissipation of the field. The results of this paper will be applied to full magneto-thermal simulations of the crust in spherical geometry in a later paper.

In Section~\ref{sec:SimEquations} we outline the form of the magnetic induction equation for electron MHD with Landau quantization-induced dHvA oscillations of the magnetization and its partial derivatives. Section~\ref{sec:SimSetup} describes the numerical method and initial conditions for the simulations, and reviews their limitations. Section~\ref{sec:SimResults} provides the simulation results, first those for simple Ohmic decay of the field including magnetization terms, and then including both Ohmic and Hall terms. Comparisons to analogous simulations neglecting the Landau quantization-induced magnetization are performed to show the significance of this effect. The implications of these results on neutron star evolution are discussed in Section~\ref{sec:Conclusion}. We work in Gaussian units throughout.

\section{Electron MHD with Landau Quantized Electrons}
\label{sec:SimEquations}

We consider the magnetic field evolution of neutron star crust matter at finite temperature. We focus on the field evolution in the crust for two reasons: 1) the electrical conductivity is much lower there than in the core, and Ohmic dissipation of the field and the heat generated by this will be concentrated in the crust; 2) Landau quantization effects are more prominent at the lower densities typical of the crust than those of the core. Since the goal is to study novel magnetization-related effects, we consider only nonrelativistic physics, which is sufficient as a first approximation to determine how important these effects are and whether it is necessary to consider them in more complicated, relativistic simulations.

As discussed in e.g., RW23, in a neutron star crust, the electric current is due to the electrons moving relative to the approximately stationary, neutralizing nuclear lattice. Under these conditions, the Hall term must be included in Ohm's Law, and the corresponding magnetohydrodynamics formulation is called electron MHD~\citep{Lighthill1960,Kingsep1990,Gordeev1994}; if the nuclear lattice can move, the term Hall MHD is used. Assuming isotropic conductivity, and ignoring the electron inertia, the electron pressure gradient and thermoelectric terms, Ohm's Law thus takes the form
\begin{equation}
\bm{E}'=\bm{E}+\frac{1}{c}\bm{v}\times\bm{B}=\frac{1}{n_{\rm e}ec}\bm{J}\times\bm{B}+\frac{1}{\sigma}\bm{J}.
\label{eq:OhmsLaw}
\end{equation}
$\bm{E}'$ is the electric field in the rest frame of the crustal lattice, which moves with velocity $\bm{v}$, $\bm{E}$ is the electric field in the frame where the crustal lattice moves with velocity $\bm{v}$, $\bm{B}$ is the magnetic field, and $\bm{J}$ is the (free) electric current density. $n_{\rm e}$ is the electron number density and $e>0$ the charge of the proton. $\sigma$ is the electrical conductivity, which  is assumed isotropic in this work, although such an approximation is generally not valid for strong magnetic fields (see, e.g.,~\citet{Potekhin1999a}). We will treat $\sigma$ as a free parameter, choosing values for it that are reasonable within a neutron star crust, and will leave consideration of the full $B$-dependent form of the anisotropic conductivity of electrons with quantization effects for a subsequent paper.

The (free) current density follows from Amp\`{e}re's Law
\begin{equation}
\bm{J}=e(Zn_{\text{N}}\bm{v}-n_{\rm e}\bm{v}_{\rm e})=\frac{c}{4\pi}\bm{\nabla}\times\bm{H},
\label{eq:AmperesLaw}
\end{equation}
where $Z$ is the atomic number of the nuclei and $n_{\text{N}}$ is the number density of nuclei. We assume local charge neutrality $Zn_{\text{N}}=n_{\rm e}$. $\bm{H}=\bm{B}-4\pi\bm{M}$ where $\bm{M}$ is the magnetization field. $\bm{M}$ is computed from the grand potential density for the magnetized electrons $\Omega_{\rm e}(B,\mue,T)$, excluding the vacuum magnetic field contribution $B^2/(8\pi)$:
\begin{equation}
\bm{M}=-\left.\frac{\partial \Omega_{\rm e}}{\partial B}\right|_{T,\mu_{\rm e}}\bm{\hat{B}}.
\end{equation}
where temperature $T$ and electron chemical potential $\mu_{\rm e}$ are held fixed. $B$ and $\hat{\bm{B}}$ are the magnitude and direction vector of $\bm{B}$, respectively. When needed, the electron number density is also computed from $\Omega_{\rm e}$:
\begin{equation}
n_{\rm e}=-\left.\frac{\partial \Omega_{\rm e}}{\partial \mu_{\rm e}}\right|_{T,B}.
\end{equation}

We will only consider the case where the nuclear lattice is fixed $\bm{v}=0$. The magnetic induction equation then becomes
\begin{equation}
\frac{\partial\bm{B}}{\partial t}=-\bm{\nabla}\times\left(\frac{1}{n_{\rm e}e}\bm{J}\times\bm{B}\right)-c\bm{\nabla}\times\left(\frac{\bm{J}}{\sigma}\right).
\label{eq:MagneticInduction}
\end{equation}
For the sake of computational convenience, we work with the magnetic vector potential $\bm{A}$ defined through $\bm{B}=\bm{\nabla}\times\bm{A}$ instead of $\bm{B}$ directly, and impose the Coulomb gauge $\bm{\nabla}\cdot\bm{A}=0$. This automatically guarantees that $\bm{B}$ remains divergence-free. Eq.~(\ref{eq:MagneticInduction}) is hence replaced by
\begin{equation}
\frac{\partial\bm{A}}{\partial t}=-c\bm{\nabla}\Phi-\frac{1}{n_{\rm e} e}\bm{J}\times\bm{B}-c\frac{\bm{J}}{\sigma},
\label{eq:MagneticInductionA}
\end{equation}
where we obtain the extra gradient of a scalar electric potential $\Phi$, which has no physical consequences for the magnetic induction equation, but which will be used in the numerical evolution of $\bm{A}$. We can write Eq.~(\ref{eq:AmperesLaw}) as
\begin{equation}
\bm{J}=\frac{c}{4\pi}\left(1-4\pi\frac{M}{B}\right)\bm{\nabla}\times\bm{B}-c\left(\bm{\nabla}M-\frac{M}{B}\bm{\nabla}B\right)\times\bm{\hat{B}}.
\label{eq:JeExpansion1}
\end{equation}
To make explicit the contributions from the (differential) magnetic susceptibility $\chi_{\mu}$ and the two mixed partial derivatives $\mathcal{M}_{\mu}$ and $\mathcal{M}_T$, defined as
\begin{align}
\chi_{\mu}\equiv\left.\frac{\partial M}{\partial B}\right|_{T,\mu_{\rm e}}, && \mathcal{M}_{\mu}\equiv\left.\frac{\partial M}{\partial \mu_{\rm e}}\right|_{T,B}, &&
\mathcal{M}_T\equiv\left.\frac{\partial M}{\partial T}\right|_{\mu_{\rm e},B}, 
\label{eq:MagnetizationPartialDerivatives}
\end{align}
we write
\begin{equation}
\bm{\nabla}M = \chi_{\mu}\bm{\nabla}B+\mathcal{M}_{\mu}\bm{\nabla}\mu_{\rm e}+\mathcal{M}_T\bm{\nabla}T.
\end{equation}
So Eq.~(\ref{eq:JeExpansion1}) can be simplified to
\begin{align}
\bm{J}={}&\frac{c}{4\pi}\left(1-4\pi\frac{M}{B}\right)\bm{\nabla}\times\bm{B}
\nonumber
\\
{}&-c\left[\left(\chi_{\mu}-\frac{M}{B}\right)\bm{\nabla}B+\mathcal{M}_{\mu}\bm{\nabla}\mu_{\rm e}+\mathcal{M}_T\bm{\nabla}T\right]\times\bm{\hat{B}}.
\label{eq:JMagnetization}
\end{align}
$\bm{B}$, $B$ and $\bm{\hat{B}}$ are then computed using $\bm{A}$. This procedure, which explicitly includes $\chi_{\mu}$, $\mathcal{M}_{\mu}$ and $\mathcal{M}_T$ in the evolution equations, is required if we want to study the effect of the large-amplitude dHvA oscillations of $\chi_{\mu}$ without using extremely fine spatial resolution in our numerical implementation. Otherwise, the finite differencing of $M$ would not capture the large oscillations of $M$ and its derivatives. Since we assume uniform temperature in our simulations, the $\mathcal{M}_T\bm{\nabla}T$ term is absent from our simulations, and we exclude it from equations in the remainder of this paper.

A more intuitive way to understand the how the magnetization and its derivatives affect the field evolution can be shown by a slight manipulation of Eq.~(\ref{eq:JMagnetization}). Assume for simplicity that $\mue$ and $T$ are uniform. Define the components of the $H\rightarrow B$ current density parallel and perpendicular to the magnetic field,
\begin{align}
\bm{J}_{B,\parallel}={}&\frac{c}{4\pi}\left(\hat{\bm{B}}\cdot\bm{\nabla}\times\bm{B}\right)\hat{\bm{B}},
\\
\bm{J}_{B,\perp}={}&\frac{c}{4\pi}\bm{\nabla}\times\bm{B}-\bm{J}_{B,\parallel}
\nonumber
\\
={}&\frac{c}{4\pi}\left[\bm{\nabla}B\times\hat{\bm{B}}+B\bm{\nabla}\times\hat{\bm{B}}\cdot\left(\bm{I}-\hat{\bm{B}}\otimes\hat{\bm{B}}\right)\right],
\end{align}
where $\bm{I}$ is the identity tensor. Using that $4\pi M/B\ll 1$, Eq.~(\ref{eq:JMagnetization}) can then be rewritten as
\begin{align}
\bm{J}\approx {}&\bm{J}_{B,\parallel}+\left(1-4\pi\chi_{\mu}\right)\bm{J}_{B,\perp}
+c\chi_{\mu}B\bm{\nabla}\times\hat{\bm{B}}\cdot\left(\bm{I}-\hat{\bm{B}}\otimes\hat{\bm{B}}\right).
\label{eq:JMagnetizationSplit}
\end{align}
Ignoring the final term, we see that the perpendicular $H\rightarrow B$ current density is increased by a factor $1-4\pi\chi_{\mu}$, which means that the perpendicular conductivity is decreased by this factor. Since $\bm{J}_{B,\parallel}\times\bm{B}=0$, the Hall term prefactor will be increased by a factor $1-4\pi\chi_{\mu}$. As we show in Section~\ref{sec:Microphysics}, the factor $1-4\pi\chi_{\mu}$ may be as great as $\sim 5$, both enhancing the dissipation rate of the field and increasing the rate of field advection by the Hall term, which further generates small-scale field components and in turn increases field dissipation further.

Based on Eq.~(\ref{eq:JMagnetization}) we can make predictions about which conditions will support the greatest enhancement to the Ohmic decay of the field due to Landau quantization-induced magnetization. Because the terms proportional to $\chi_{\mu}$ and $\mathcal{M}_{\mu}$ scale as $\bm{\nabla}B$ and $\mue$ respectively, the enhancement of the Ohmic decay is greatest where the field and $\mue$ gradients are largest. In these regions, the number of occupied Landau levels can change rapidly, assuming that the field is not so strong as to confine the electrons to a small number of Landau levels. For field and number densities such that the occupied number of Landau levels is $\lesssim 5$, the dHvA oscillations in $\chi_{\mu}$ are insufficiently large to lead to significant enhancement of Ohmic decay (see Figure~\ref{fig:MChiM}, third panel).

\subsection{Energy conservation}
\label{sec:EnergyConservation}

To determine whether MHD simulations are numerically stable, energy conservation is monitored. In the case of $\bm{H}\neq\bm{B}$, the magnetic energy conservation condition is derived by taking the dot product of Eq.~(\ref{eq:MagneticInduction}) with $\bm{H}$, giving
\begin{equation}
\frac{1}{4\pi}\bm{H}\cdot\frac{\partial\bm{B}}{\partial t} = -\bm{\nabla}\cdot\bm{S}-\frac{1}{\sigma}J^2,
\end{equation}
where
\begin{equation}
\bm{S}=\frac{c}{4\pi}\bm{E}\times\bm{H},
\end{equation}
is the Poynting vector and the left-hand side can be rewritten using the first law of thermodynamics as~\citep{Landau1960}
\begin{equation}
\bm{H}\cdot\frac{\partial\bm{B}}{\partial t} = 4\pi\left(\frac{\partial \Omega}{\partial t}+s\frac{\partial T}{\partial t}+n\frac{\partial \mu}{\partial t}\right),
\label{eq:EnergyConservation1}
\end{equation}
where $\Omega=\Omega(\mu,T,\bm{B})$ is the total grand potential density. $s$, $T$, $\mu$ and $n$ are the entropy density, temperature, chemical potential and number density of the medium. $\Omega$ includes contributions from both the electrons and the nucleons in the crust and the magnetic field. Eq.~(\ref{eq:EnergyConservation1}) says there is no way to separate the magnetic and matter energy contributions in the energy conservation equation, and in order to monitor energy conservation using this equation, we would need to compute $\Omega$, $\mu$ and $n$.

Instead of considering Eq.~(\ref{eq:EnergyConservation1}), we monitor conservation of a ``quasi-energy'' derived by taking the dot product of Eq.~(\ref{eq:MagneticInduction}) with $\bm{B}/(4\pi)$, giving
\begin{equation}
\frac{\partial}{\partial t}\left(\frac{B^2}{8\pi}\right)=-\bm{\nabla}\cdot\bm{S}_B-\frac{1}{\sigma}\bm{J}_B\cdot\bm{J}-\frac{1}{n_{\rm e}ec}\bm{J}_B\cdot\left(\bm{J}_{\nabla}\times\bm{B}\right),
\label{eq:QuasiEnergyConservation}
\end{equation}
where
\begin{subequations}
\begin{align}
\bm{J}_B\equiv{}&\frac{c}{4\pi}\bm{\nabla}\times\bm{B},
\\
\bm{S}_B\equiv{}&\frac{c}{4\pi}\bm{E}\times\bm{B},
\\
\bm{J}_{\nabla}\equiv{}&-4\pi\frac{M}{B}\bm{\nabla}\times\bm{B}-4\pi\left[\left(\chi_{\mu}-\frac{M}{B}\right)\bm{\nabla}B+\mathcal{M}_{\mu}\bm{\nabla}\mu_{\rm e}\right]\times\bm{\hat{B}}.
\label{eq:Jgrad}
\end{align}
\end{subequations}
Unlike in Eq.~(\ref{eq:EnergyConservation1}), we are unable to entirely eliminate the Hall term from the quasi-energy conservation equation, and hence the presence of the term containing $\bm{J}_{\nabla}$ in Eq.~(\ref{eq:QuasiEnergyConservation}) which vanishes in the $\bm{H}=\bm{B}$ limit. Volume integrating Eq.~(\ref{eq:QuasiEnergyConservation}) over the simulation domain gives
\begin{align}
\frac{\text{d}U_B}{\text{d}t} ={}& -\oint\bm{S}_B\cdot\hat{\bm{n}}\text{d}^2x
\nonumber
\\
{}&-\left[\int\bm{J}_B\cdot\left(\frac{\bm{J}}{\sigma}+\frac{1}{n_{\rm e}ec}\left(\bm{J}_{\nabla}\times\bm{B}\right)\right)\text{d}^3x\right],
\label{eq:QuasiEnergyConservationIntegrated}
\end{align}
where $\hat{\bm{n}}$ is a unit normal vector to the integration volume and
\begin{equation}
U_B = \frac{1}{8\pi}\int B^2\text{d}^3x.
\end{equation}
We monitor the energy balance by comparing the change in $U_B$ as a function of time with the time integral of the right-hand side of Eq.~(\ref{eq:QuasiEnergyConservationIntegrated}). Since we consider a fully periodic box for our simulation, there are no Poynting flux losses out of the domain, and the first term on the right-hand side is zero.

\begin{figure}
\center
\includegraphics[width=0.98\linewidth]{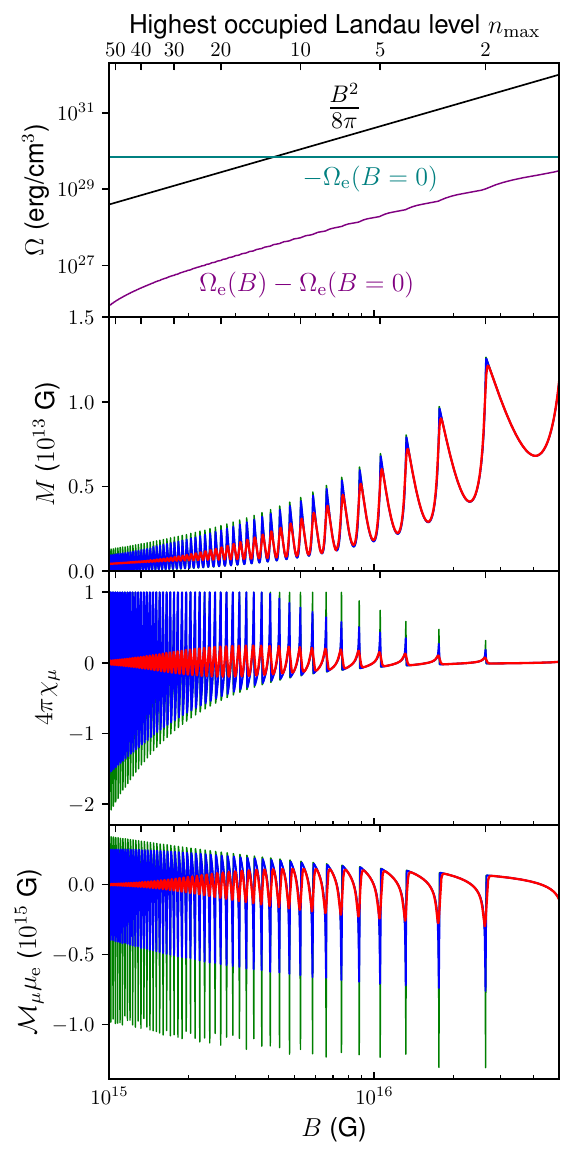}
\caption{Contributions to $\Omega$ (top), $M$, $4\pi\chi_{\mu}$ and $\mathcal{M}_{\mu}$ times $\mue$ as a function of $B$ for fixed $\mue=25$ MeV. In the top panel, $T=2\times10^8$ K is used; in the bottom three panels, three different temperatures are used: $T=10^{9}$ K (red, thickest), $2\times10^{8}$ K (blue, intermediate) and $7\times10^7$ K (green, thin). The upper horizontal axis shows the number of occupied Landau levels at $T=0$, determined from $n_{\rm max}=\lfloor(\mue^2-m_e^2)/(2eB)\rfloor$. The de Haas--van Alphen oscillations in $M$, $4\pi\chi_{\mu}$ and $\mathcal{M}_{\mu}$ are clearly visible, and are less pronounced but still visible in $\Omega_{\rm e}(B)-\Omega_{\rm e}(B=0)$. Thermodynamic stability has been enforced on $\chi_{\mu}$ such that $\chi_{\mu}\leq 1/(4\pi)$.}
\label{fig:MChiM}
\end{figure}

We can justify why we only consider the conservation of the ``quasi-energy''. Ignoring the magnetic interaction with the nucleons, the grand potential density of the magnetized crust is
\begin{equation}
\Omega = \Omega_{\rm e}(\mue,T,B) + \frac{B^2}{8\pi} + \Omega_{\rm N}(\mu_{\rm N},T),
\end{equation}
where $\Omega_{\rm N}$ is the contribution of the nucleons that have chemical potential $\mu_{\rm N}$. For a fixed temperature and $\mu_{\rm N}$, we then have
\begin{equation}
\frac{\partial\Omega}{\partial t} = \frac{\partial}{\partial t}\left(\frac{B^2}{8\pi}\right) + \frac{\partial\Omega_{\rm e}}{\partial t},
\end{equation}
If the $B$-independent part of $\Omega_{\rm e}$, which does not change for fixed $\mu_{\rm e}$ and $T$, is subtracted off, then the remaining contribution is two orders of magnitude smaller than $\frac{B^2}{8\pi}$, as is shown in the top panel of Figure~\ref{fig:MChiM}. Hence, the vast majority of the energy dissipated in our fixed $T$ simulations will be the ``quasi-energy'' $U_B$, and monitoring its conservation is sufficient to ensure total energy conservation to within a few percent.

\subsection{Microphysics of neutron star crust matter}
\label{sec:Microphysics}

As discussed in the opening paragraph, strong magnetic fields and the resulting Landau quantization of electrons have long been known to modify the thermodynamic and transport properties of matter. The main goal of this work is to include these modifications to the magnetization $M$, differential magnetic susceptibility $\chi_{\mu}$ and the mixed partial derivative $\mathcal{M}_{\mu}$ in magnetic field evolution simulations for the first time. For simplicity, we \textit{do not} include the effects of Landau quantization on transport coefficients in this work, which allows us to isolate the effect of the dHvA oscillations of $M$, $\chi_{\mu}$ and $\mathcal{M}_{\mu}$ on the field evolution.

$M$, $\chi_{\mu}$ and $\mathcal{M}_{\mu}$ are functions of $B$, $\mue$ and $T$ and are computed in the Appendix. The Appendix describes the two regimes of approximation, depending on the temperature, used in the numerical calculations to avoid performing costly Fermi--Dirac integrals at each time step. As discussed in Appendix~\ref{app:HighTApprox}, the effect of temperature on $M$, $\chi_{\mu}$ and $\mathcal{M}_{\mu}$ is determined through the ratio $2\pi^2\mue T/(eB)$. If this ratio is large, i.e., $T\gg eB/(2\pi^2\mue)$, the dHvA oscillations are thermally suppressed, while if $T\ll eB/(2\pi^2\mue)$ thermal effects are relatively unimportant. We thus use different numerical approximations for $T>eB/(2\pi^2\mue)$ and $T\leq eB/(2\pi^2\mue)$.

We choose temperature values $5\times10^7\text{ K}<T<1\times10^9\text{ K}$. This includes the expected temperatures for neutron star crusts with ages between $\sim 100$ yr and $\sim 100$ kyr, and extends to lower temperatures where the effects of Landau quantization should be more obvious. We assume uniform and fixed temperature over the simulation domain. Because it controls the amplitude of the dHvA oscillations, temperature is important in determining whether the enhanced Ohmic dissipation effect due to said oscillations is active or not. We choose artificial $\mue$ profiles with values typical of the inner crust of a neutron star, $25\text{ MeV}\lesssim\mue\lesssim 80\text{ MeV}$. The electron number density $n_{\rm e}=n_{\rm e}(B,\mue,T)$ is featured in the magnetic induction equation in the Hall term, and we compute it at each time step. The difference between $n_{\rm e}(B)$ and $n_{\rm e}(B=0)$ is small except in the strongly quantized limit, so $n_{\rm e}$ does not change significantly during the evolution. Like the temperature, we use uniform electrical conductivities appropriate for neutron star inner crusts $10^{22}\text{ s}^{-1}\lesssim\sigma\lesssim 10^{24}\text{ s}^{-1}$, ignoring the temperature and density-dependence of $\sigma$. We fix the value $\sigma = 5\times10^{22}\text{ s}^{-1}$ for all of our simulations.

Previous papers~\citep{Blandford1982,Suh2010,Wang2013,Wang2016,Rau2023} have considered the formation of magnetic domains within a strongly magnetized neutron star crust. These domains form in response to a thermodynamic instability when the differential magnetic susceptibility $\chi_{\mu}$ is greater than $1/(4\pi)$, and in equilibrium, in these unstable regions of parameter space, $\chi_{\mu}$ will take a value slightly less than $1/(4\pi)$. The exact value can be determined using a Maxwell construction procedure as described in e.g. RW23. 

We make the assumption that the magnetic domains form on timescales much shorter than the timescales of large-scale field evolution. If the domains form on the Ohmic timescale
\begin{equation}
\tau_{\rm O}=\frac{4\pi L^2\sigma}{c^2},
\label{OhmicTimescale}
\end{equation}
for domain length scale $L\sim 1$ cm--$1$ m, which is at least $3$ orders of magnitude shorter than the thickness of a neutron star crust $L~\sim 1$ km, then the domain timescale is $\lesssim 6$ orders of magnitude shorter than the large-scale field Ohmic timescale. The Hall timescale is
\begin{equation}
\tau_{\rm H}=\frac{4\pi en_{\rm e} L^2}{cB}.
\label{eq:HallTimescale}
\end{equation}
The ratio of the Hall to Ohmic timescales is independent of length scale, and for magnetar-strength fields is
\begin{equation}
\frac{\tau_{\rm H}}{\tau_{\rm O}}=\frac{n_{\rm e}ec}{B\sigma}=1.4\times10^{-2}\left(\frac{n_{\rm e}}{10^{-4}\textrm{ fm}^{-3}}\right)\left(\frac{10^{15}\textrm{ G}}{B}\right)\left(\frac{10^{23}\textrm{ s}^{-1}}{\sigma}\right),
\end{equation} 
so the domain formation timescale is $\lesssim 4$ orders of magnitude shorter than the large-scale field Hall timescale.

Determining the exact size and location of the magnetic domains using the Maxwell construction (see Section~5.2 of RW23) would be numerically prohibitive, since for fields with large gradients required for the term $\propto\chi_{\mu}$ in Eq.~(\ref{eq:JMagnetization}) to be significant, tens to hundreds of domains are expected to be formed within the simulation volume. We instead use the approximate criterion that domains will form where $\chi_{\mu}>1/(4\pi)$, and so set $\chi_{\mu}=1/(4\pi)$ when this condition is true such that $\partial^2\Omega/\partial B^2=(1/(4\pi)-\chi_{\mu})\geq 0$ and thermodynamic stability always holds. Similarly, the stability condition for $\mathcal{M}_{\mu}$ is $\mathcal{M}_{\mu}^2>(1/(4\pi)-\chi_{\mu})\Omega_{\mu\mu}$, where $\Omega_{\mu\mu}=\partial^2\Omega_{\rm e}/\partial \mue^2|_B<0$. So if $\chi_{\mu}\leq 1/(4\pi)$ is imposed, this condition is trivially satisfied. 

Figure~\ref{fig:MChiM} shows $M$, $\chi_{\mu}$ and $\mathcal{M}_{\mu}$ as a function of $B$ for fixed $\mue$ and different temperatures to illustrate the typical values for these parameters. An important quantity is the maximum occupied Landau level at $T=0$, given by
\begin{equation}
n_{\rm max}=\left\lfloor\frac{\mue^2-m_{\rm e}^2}{2eB}\right\rfloor.
\label{eq:nmax}
\end{equation}
The peaks of the dHvA oscillations occur when a new Landau level becomes empty/occupied, i.e., whenever $(\mue^2-m_{\rm e}^2)/(2eB)$ takes exact positive integer values. At $T=0$, $\chi_{\mu}$ and $\mathcal{M}_{\mu}$ are divergent at such values; these divergences are suppressed by finite $T$. We can use this to determine the approximate length scale between the crests of the dHvA oscillations: given a magnetic field with characteristic length scale $\ell_B=|\bm{\nabla}B/B|^{-1}$, the distance between the crests should be roughly
\begin{equation}
\ell_{\rm{dHvA}}=\left|\bm{\nabla}\left\lfloor\frac{(\mue^2-m_{\rm e}^2)}{2eB}\right\rfloor\right|^{-1}=\frac{\ell_B}{n_{\rm max}},
\label{eq:dHvALengthScale}
\end{equation}
where we assume uniform $\mue$. If the length scale of the field is the order of the crust thickness $\ell_B\sim 1$ km, then the length scale between neighboring dHvA crests for the conditions shown in Figure~\ref{fig:MChiM} would be $\ell_{\rm{dHvA}}\geq 20$ m. We will examine fields as low as $10^{14}$ G in our numerical simulations, and so length scales an order of magnitude lower than this would be reasonable in that case. Additionally, this is simply the length scale between oscillation crests, so a finer spatial resolution than this would be needed to resolve the oscillation itself. As the field evolves, $\ell_B$ can also become shorter locally, further reducing $\ell_{\rm{dHvA}}$.

\section{Numerical Simulations: Setup}
\label{sec:SimSetup}

We simulate the evolution of the magnetic field via the vector potential $\bm{A}$ within a periodic box of neutron star crust matter in Cartesian coordinates. The initial field configuration only depends on the $y$-and $z$-coordinates, and $\mue$ is allowed to depend on the $z$-coordinate only. This restricts the allowed form of the initial vector potential to periodic functions in the $y$ and $z$-directions. We work in the Coulomb gauge $\bm{\nabla}\cdot\bm{A}=0$, further restricting $\bm{A}$. We thus take as the form of $\bm{A}$
\begin{subequations}
\begin{align}
A_x(y,z)={}& B_m\left[ \frac{L_y}{2\pi}\cos\left(\frac{2\pi y}{L_y}\right) + \beta\frac{L_z}{2\pi p_B}\sin\left(\frac{2\pi z p_B}{L_z}\right) \right],
\\
A_y(z)={}& \beta B_m\frac{L_z}{2\pi p_B}\sin\left(\frac{2\pi z p_B}{L_z}\right),
\\
A_z(y)={}& B_m\frac{L_y}{2\pi}\cos\left(\frac{2\pi y}{L_y}\right),
\end{align}
\end{subequations}
where the dimensions of the periodic box are $L_x\times L_y\times L_z$, $B_m$ sets the field magnitude and $\beta\geq 0$ is a dimensionless parameter allowing the relative strengths of the components of the field to be varied. $p_B\geq 1$ is an integer chosen to generate additional oscillations of the magnetic field in the $z$-direction. For $\mue$ we choose the profile
\begin{equation}
\mue(z) = \mue\left[1 + \nu\cos\left(\frac{2\pi p_n z}{L_z}\right)\right],
\end{equation}
where $n_{\rm e, 0}$ is the characteristic electron number density, $\nu$ is the amplitude of the fluctuations in $\mue$ and $p_n$ is a positive integer. We fix $p_n=3$ and $L_x=L_y=L_z=1$ km throughout our simulations. This form of the field allows us to demonstrate the enhancement of magnetic field dissipation due to the Landau quantization-induced differential magnetic susceptibility and how this effect changes with $B$, $\mue$ and $T$. The choices of initial field and $\mue$ configurations that we use are listed in Tables~\ref{tab:BConfigs} and~\ref{tab:neConfigs} respectively. We emphasize that most simulations will consider the uniform $\mue$ profile $\upmu1$. Additionally, Table~\ref{tab:BConfigs} shows the range of $n_{\rm max}$ for a given initial field configuration, assuming $\mue$ configuration $\upmu1$ in Table~\ref{tab:neConfigs}. The range in $n_{\rm max}$ values is a measure of the number of dHvA oscillations of the magnetization and its partial derivatives that occur within the simulation domain, with a wider range indicating that more dHvA oscillations will occur.

\begin{table}
	\caption{Parameter configurations for the magnetic field.}
	  \centering
    \begin{tabular}{|c|c|c|c|c|c|}
    \hline
	\multicolumn{1}{|c|}{Name} & $B_m$ (G) & $p_B$ & $\beta$ & $n_{\rm max,low}$ & $n_{\rm max,high}$\\ 
    \hline
    	A1 & $5\times 10^{14}$ & $1$ & $10$ & $13$ & $279$ \\
	\hline  
	    A2 & $2.5\times 10^{14}$ & $1$ & $10$ & $27$ & $559$ \\
	\hline 
	    A3 & $5\times 10^{14}$ & $2$ & $10$ & $13$ & $279$ \\
	\hline
		A4 & $1\times 10^{14}$ & $2$ & $10$ & $69$ & $1399$ \\
	\hline
		A5 & $5\times 10^{14}$ & $1$ & $1$ & $92$ & $292$\\
	\hline
		A6 & $5\times 10^{15}$ & $1$ & $10$ & $1$ & $29$\\
	\hline
    \end{tabular}
    \label{tab:BConfigs}
\end{table}

\begin{table}
	\caption{Parameter configurations for the electron chemical potential profile.}
	  \centering
    \begin{tabular}{|c|c|c|}
    \hline
	\multicolumn{1}{|c|}{Name} & $\mu_{\rm e}$ (MeV) & $\nu$\\ 
    \hline
    	$\upmu$1 & $35$ & $0$\\
	\hline  
	    $\upmu$2 & $35$ & $0.1$\\
	\hline  
	    $\upmu$3 & $35$ & $0.2$\\
	\hline 
    \end{tabular}
    \label{tab:neConfigs}
\end{table}

To perform the simulations, we used \texttt{Dedalus v.3}~\citep{Burns2020}, a partial differential equation-solving package for Python employing spectral methods. \texttt{Dedalus} uses symbolic equation entry, which allowed us to include nonstandard physics such as the differential magnetic susceptibility terms easily within our evolution equations. \texttt{Dedalus} performs time stepping of linear terms implicitly and nonlinear terms explicitly. This partially avoids the Courant--Friedrich--Lewy condition constraint on the maximum time step/minimum grid spacing for stability. Numerical instability cannot be entirely avoided if nonlinear terms are present, as they are for the problem consider here, so the maximum time step must be chosen carefully.

To solve the magnetic induction equation Eq.~(\ref{eq:MagneticInductionA}) numerically, the Ohmic term is split into parts that are linear and nonlinear in the vector potential: 
\begin{equation}
\frac{\partial\bm{A}}{\partial t} - \frac{c^2}{4\pi\overline{\sigma}}\nabla^2\bm{A} + c\bm{\nabla}\Phi = \frac{c^2}{4\pi\tilde{\sigma}}\nabla^2\bm{A} -\frac{c}{\sigma}\bm{J}_{\nabla}-\frac{c}{4\pi n_{\rm e}e}\bm{J}\times\bm{B}.
\label{eq:MagneticInductionASplit}
\end{equation}
The left-hand side terms are linear in $\bm{A}$ and are time stepped implicitly, and the right-hand side terms are nonlinear in $\bm{A}$ and time stepped explicitly. $\overline{\sigma}$ is a constant, uniform conductivity and $\tilde{\sigma}$ represents fluctuations away from this value such that $\sigma^{-1}=\overline{\sigma}^{-1}+\tilde{\sigma}^{-1}$. $\tilde{\sigma}$ can also depend on $B$ if we were e.g., including the effect of the magnetic field on transport. Since we only consider uniform $\sigma$, this splitting is not strictly necessary, but choosing $\overline{\sigma}$ appropriately can increase numerical stability as discussed below.

Eq.~(\ref{eq:MagneticInductionASplit}) is made dimensionless by dividing by characteristic length scale $L_0$, field strength $B_0$, electron density $n_{{\rm e}0}$ and time scale $\tau$ where appropriate, giving
\begin{align}
\frac{\partial\overline{\bm{A}}}{\partial\overline{t}} - \overline{\eta}_{\rm O}\overline{\nabla}^2\bm{\overline{A}} + \overline{\bm{\nabla}}\ \overline{\Phi} = \tilde{\eta}_{\rm O}\overline{\nabla}^2\overline{\bm{A}} -\eta_{\rm O}\bm{j}_{\nabla}-\eta_{\rm H}\bm{j}\times\bm{\overline{B}},
\label{eq:MagneticInductionASplitDedimensionalized}
\end{align}
where $\bm{\overline{A}}=\bm{A}L_0/B_0$, $\overline{\bm{\nabla}}=L_0\bm{\nabla}$, etc. $\eta_{\rm O}$ and $\eta_{\rm H}$ are the dimenionless Ohmic and Hall diffusivities
\begin{align}
\eta_{\rm O}\equiv{}&\frac{c^2\tau}{4\pi\sigma L_0^2},
\label{eq:OhmicDiffusivity}
\\
\eta_{\rm H}\equiv{}&\frac{c\tau B_0}{4\pi n_{\rm e}e L_0^2}.
\label{eq:HallDiffusivity}
\end{align}
The constant and fluctuating electrical conductivities are absorbed into the dimensionless Ohmic diffusivities $\overline{\eta}_{\rm O}$ and $\tilde{\eta}_{\rm O}$, respectively. The value of $\overline{\eta}_{\rm O}$ is chosen such that the $\tilde{\eta}_{\rm O}\nabla^2\bm{A}$ term is antidiffusive, which is necessary for numerical stability. The reduced current densities are defined as
\begin{align}
\bm{j}\equiv \frac{4\pi}{c}\overline{\bm{J}}, \qquad 
\bm{j}_{\nabla}\equiv \frac{4\pi}{c}\overline{\bm{J}}_{\nabla},
\end{align}
where $\overline{\bm{J}}$ and $\overline{\bm{J}}_{\nabla}$ are the de-dimensionalized forms of Eq.~(\ref{eq:JMagnetization}) and Eq.~(\ref{eq:Jgrad}), respectively. The characteristic scales we use are $B_0=10^{15}$ G, $L_0=10^4$ cm, $n_{{\rm e}0}=10^{-4}$ fm$^{-3}$ and $\tau=4\pi en_{{\rm e}0}L_0^2/(B_0c)=2.02\times10^9$ s $=64$ yr.

There is some freedom to choose the value of $\overline{\eta}_{\rm O}=\eta_{\rm O}-\tilde{\eta}_{\rm O}$. Choosing a large value helps avoid numerical instability, since the stiff left-hand side of Eq.~(\ref{eq:MagneticInductionASplitDedimensionalized}) is treated implicitly, while the large antidiffusive value of $\tilde{\eta}_{\rm O}$ on the right-hand side stabilizes the explicit timestepping on this side of the equation. In the literature on generalized diffusion equations, of which the magnetic induction equation is an example, the use of such a constant is referred to as the stabilization method or auxiliary variable method~\citep{He2007,Li2022}. However, if $\overline{\eta}_{\rm O}$ is taken to be too large such that $\tilde{\eta}_{\rm O}\approx-\overline{\eta}_{\rm O}$, there can be an ``overdamping'' effect, which results in inaccurate simulation results. Hence, we restricted $\overline{\eta}_{\rm O}=1000$ in our simulations, which is large compared to the uniform value of $\eta_{\rm O}$ we use ($\eta_{\rm O}=0.02$ from Eq.~(\ref{eq:OhmicDiffusivity}) with $\sigma=5\times10^{22}$ s$^{-1}$), but not so large as to make $\eta_{\rm O}$ negligibly small from a numerical perspective.

Within \texttt{Dedalus}, we use the RK222 multistep implicit-explicit (IMEX) method, which we found provided a good balance between accuracy and numerical stability out of the time stepping methods provided. We use a $3/2$ dealiasing factor to ameliorate spectral aliasing from the nonlinear terms. The simulations are run in quasi-2D: $\bm{A}$ (and hence $\bm{B}$) have three components, but only vary in two directions, which we take as $y$ and $z$. We use identical spatial resolutions in these directions, varying from $128\leq N\leq 576$ where $N=L/\Delta L=L_y/\Delta L_y=L_z/\Delta L_z$. These resolutions are certainly too low to resolve all of the dHvA oscillations of the Landau quantization-induced magnetization and its partial derivatives, whose length scale is given by Eq.~(\ref{eq:dHvALengthScale}), but resolving all of these oscillations is impractical for any simulation. The simulations hence can be considered to use smoothed average values of the coefficients with dHvA oscillations.

Our simulations involve nonlinear terms--the Hall term and all terms depending on $M$, $\chi_{\mu}$ and $\mathcal{M}_{\mu}$--and highly oscillatory, spatially varying coefficients. The Hall term sets the limit on the time step for a given grid spacing: anything that decreases the Hall time Eq.~(\ref{eq:HallTimescale}) (increased $B$, decreased $n_{\rm e}$, increased spatial resolution) will reduce the time step necessary for stability. The Hall timescale increases as the magnetic field decreases, and hence the time step required for stability increases, but since the magnetic field decreases on the slower Ohmic decay timescale, we use constant, resolution-dependent time steps for the magnetic evolution simulations. The inclusion of Landau quantization-induced magnetization terms necessitates a somewhat smaller time step than including the Hall term alone, with time steps about a factor of 1.5-2 smaller required for stability and energy conservation with Landau quantization \textit{and} the Hall term than with the latter alone. Additionally, computing coefficients $M$, $\chi_{\mu}$ and $\mathcal{M}_{\mu}$ that are functions of the evolving $B$ increases the run time by about an order of magnitude compared to simulations without Landau quantization effects. These restrictions mean that we did not run our simulations for the typical durations of tens of thousands of years used in many other magnetic field or magneto-thermal evolution calculations, nor were we able to use very high spatial resolutions. This means that comparison \textit{between} the simulations, particularly those with identical initial conditions with Landau quantization effects turned off and on, is most important to distinguish the effects of the novel physics. We also compare simulations with differing resolution to determine if and at which resolution numerical convergence is achieved. As we demonstrate in Section~\ref{sec:HallOhmSimulations}, the resolution required for numerical convergence can vary between initial field configurations, temperature and background $\mue$ profile, as all of these factors affect the amplitude and spatial period of the dHvA oscillations.

Energy conservation was monitored throughout the simulations according to the prescription in Section~\ref{sec:EnergyConservation}. For simulations without Landau quantization-induced magnetization terms included, energy was conserved to better than 1\%. The inclusion of Landau quantization effects worsened the energy conservation, but for those simulations, energy was conserved to within 5\%. 

\section{Numerical Simulations: Results}
\label{sec:SimResults}

\subsection{Pure Ohmic dissipation}

We first study the field evolution with the Hall term turned off. This allows us to clearly demonstrate the enhanced Ohmic dissipation associated with the Landau quantization-induced magnetization without having to worry about the possible numerical instability arising from the Hall term. However, since the Hall term itself enhances the dissipation through the Hall cascade to smaller length scales, we expect that the full impact of the enhancement of Ohmic dissipation due to dHvA oscillations will be absent without the Hall term. We denote simulations with the Landau quantization effects turned on/off as ``LQ'' or ``no LQ'' simulations, respectively.

To compare the Ohmic dissipation rate between simulations, we plot two quantities: the fractional dissipated magnetic field energy 
\begin{equation}
\frac{|\Delta U_B(t)|}{U_B(t=0)} = \frac{U_B(t=0)-U_B(t)}{U_B(t=0)},
\label{eq:FracDissME}
\end{equation}
and the Ohmic dissipation amplification factor (ODAF)
\begin{equation}
{\rm ODAF}\equiv \frac{\Delta U_B(t)}{\Delta U_B^{\rm No\ LQ}(t)},
\label{eq:ODAF}
\end{equation}
which compares the magnetic field energy dissipated with and without the Landau quantization-induced magnetization terms in the simulation at identical times. As $t\rightarrow\infty$, the ODAF will always approach 1, but for times less than the Ohmic decay time--which, for $L=1$ km and $\sigma=5\times10^{22}$ s$^{-1}$, is over $2\times10^5$ yr--it is a useful way to distinguish between the Ohmic dissipation rates of different initial configurations (magnetic field, temperature, $\mue$ profile, resolution). The No LQ simulations are temperature-independent since we set $\sigma$ as a constant, and nearly resolution-independent, so it is reasonable to compute the ODAF for a suite of simulations using a single No LQ simulation for the denominator of Eq.~(\ref{eq:ODAF}).

\begin{figure}
\center
\includegraphics[width=0.98\linewidth]{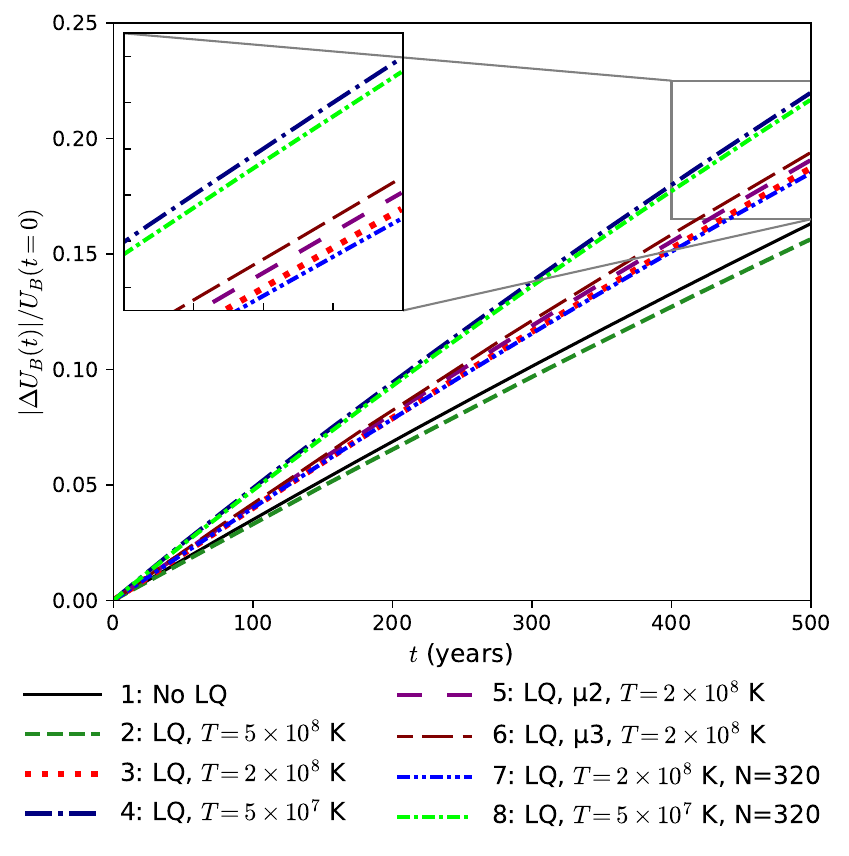}
\caption{Comparison of the fractional magnetic field energy dissipated after $t=500$ years for the A1 field configuration for seven Ohmic decay-only simulations. Fixed conductivity $\sigma=5\times 10^{22}$ s$^{-1}$ was used for all simulations, as were $\mue$ configuration $\upmu$1 and resolution $N=192$ unless otherwise specified.}
\label{fig:DeltaUBOhm}
\end{figure}

In Figure~\ref{fig:DeltaUBOhm}, we plot the fractional dissipated magnetic field energy for simulations without the Hall term at fixed $\sigma=5\times10^{22}$ s$^{-1}$. We compare a variety of temperatures, $\mue$ profiles and spatial resolutions for fixed initial field configuration A1. Comparing the No LQ simulation (1) to the first three LQ simulations (2--4), which used different temperatures, we see that as the temperature is decreased, the Ohmic dissipation increases. At $T=5\times 10^8$ K, the dissipation is nearly equal with and without LQ, but for $T=5\times10^7$ K it is increased by a factor $\sim 1.4$ compared to the No LQ case. This increase is due to the reduction in the thermal suppression of the amplitude of dHvA oscillations as illustrated in Figure~\ref{fig:MChiM}. Adding oscillations in $\mue$ by using configurations $\upmu$2 and $\upmu$3 slightly increases the Ohmic dissipation at fixed temperature (compare curve 3 to 5--6). However, this is a small effect that is much less important to changing the Ohmic dissipation rate than changing the temperature is. Finally, comparing identical simulations with different spatial resolutions (curves 3 and 7, and curves 4 and 8), we find that increasing the spatial resolution leads to a slight decrease in the Ohmic dissipation of the field. We discuss why this occurs when we examine the resolution-dependence of the simulations in further detail later in this Section.

\begin{figure}
\includegraphics[width=\columnwidth]{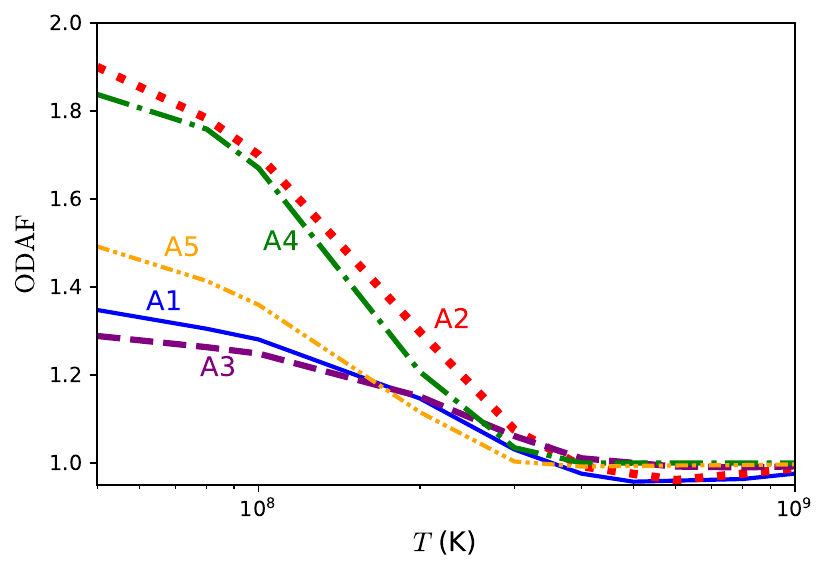}
\caption{The Ohmic dissipation amplification factor as defined in Eq.~(\ref{eq:ODAF}) as a function of temperature for Ohmic dissipation-only simulations. Five different initial magnetic field configurations are compared. A fixed conductivity $\sigma=5\times10^{22}$ $s^{-1}$, spatial resolution $N=192$, $\mue$ profile $\upmu$1 and simulation time $t=500$ yr were used for all simulations.}
\label{fig:TempCompOhm}
\end{figure}

In Figure~\ref{fig:TempCompOhm}, we plot the ODAF as a function of temperature $5\times10^7<T<10^9$ K for fixed spatial resolution with five different initial magnetic field configurations. As expected from Figure~\ref{fig:MChiM}, increasing the temperature reduces the ODAF, and in some cases this can drop below 1 at high temperatures, indicating that the inclusion of the Landau quantization effects is slightly decreasing the amount of magnetic field energy being dissipated. Configurations A2 and A4 have similar ODAF values at each temperature, and are generally larger than for configuration A1. This is because these configurations have wider ranges in $\nmax$, and hence more dHvA oscillations with higher amplitude, which generate small-scale field structure to be quickly dissipated.

\begin{figure}
\includegraphics[width=\columnwidth]{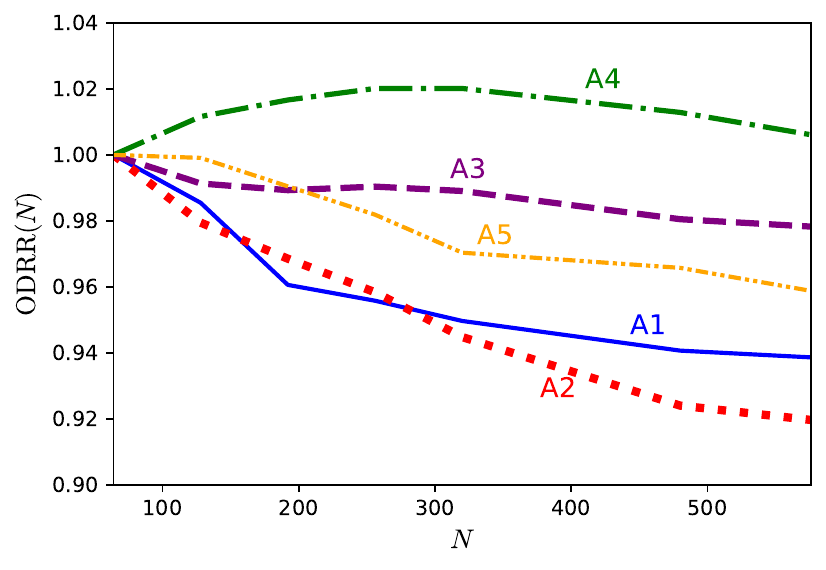}
\caption{${\rm ODRR}(N)$, the ratio of the Ohmic dissipation at resolution $N$ to the Ohmic dissipation at the lowest resolution used for a particle set of simulations ($N=64$ here), for Ohmic dissipation-only LQ simulations. Three different choices of initial field configuration were considered. Fixed conductivity $\sigma=5\times10^{22}$ $s^{-1}$, temperature $T=2\times10^8$ K, $\mue$ profile $\upmu$1 and $t=500$ yr were used for all simulations.}
\label{fig:ResolutionCompOhm}
\end{figure}

Since reasonable spatial resolutions are unable to resolve all of the dHvA oscillations, we check how the field evolution is affected by the spatial resolution $N$. In Figure~\ref{fig:ResolutionCompOhm}, we plot the Ohmic dissipation resolution ratio ${\rm ODRR}(N)$, defined as the ratio of the Ohmic dissipation at resolution $N$ to the Ohmic dissipation at the lowest resolution used for a particular set of simulations, holding all other variables fixed. ${\rm ODRR}(N)$ is by definition 1 at the lowest resolution shown. The No LQ simulations are not shown because their results are nearly resolution-independent. 

We find that the spatial resolution makes little difference in the field energy dissipated in the case of Ohmic dissipation-only simulations, though it does tend to decrease slightly as resolution is increased. This is because at lower resolutions, the spatial extent of the dHvA oscillations can be smoothed out and their effect on the field evolution is overestimated. This can decrease the effective conductivity, and hence increase the ODAF, to a greater extent than it would be if the oscillations were more accurately represented numerically. As the oscillations are more accurately resolved, this overestimation of the effective conductivity is reduced. The greater the number of dHvA oscillations within the simulation domain or the smaller their size, the more difficult it is to resolve them even as the resolution is increased, and so the ODRR should reduce more gradually as the resolution increases for initial conditions with many dHvA oscillations. This is indeed what we observe. The decrease in the ODRR with increasing resolution is most gradual for configuration A4, which increases slightly with resolution though starts to decrease by $N=576$, because this configuration has the widest initial $\nmax$ range. The initial $\nmax$ range is also broad, explaining why ODRR for it also decreases more gradually with $N$. In the case of A3, the decrease in ODRR with $N$ is more gradual than for A1 because it has shorter wavelength ($p_B=2$ compared to $p_B=1$), so the spatial extent of the dHvA oscillations is smaller than in A1 even though the initial conditions are otherwise identical.

\begin{figure}
\includegraphics[width=0.97\columnwidth]{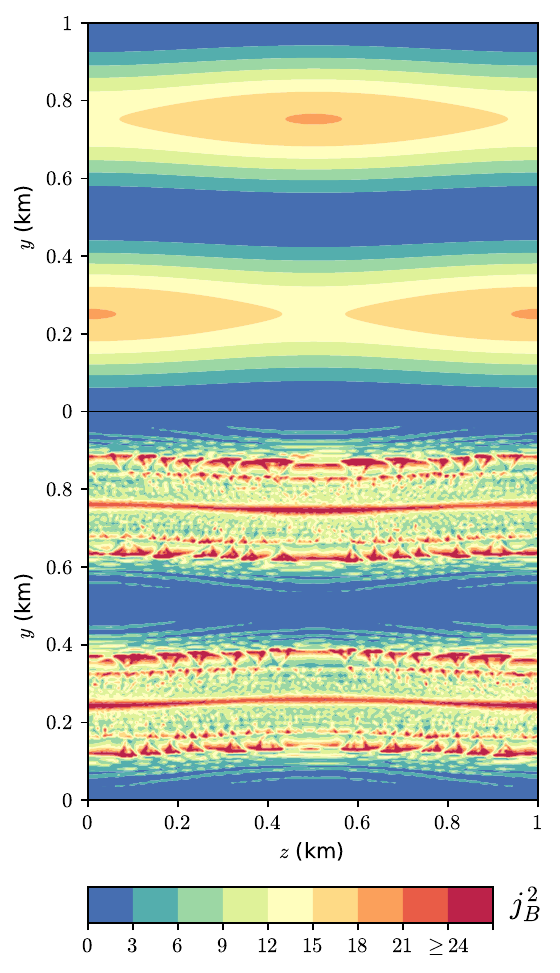}
\caption{Contour plot showing reduced current density squared $j_B^2=|\overline{\bm{\nabla}}\times\overline{\bm{B}}|^2$ at the end of 500 yr-long Ohmic decay-only simulations starting from identical initial field configuration A1. Top: Landau quantization effects turned off, $N=192$. Bottom: Landau quantization effects turned on, $N=192$. The small-scale field structure generated by this effect is clearly visible. Fixed conductivity $\sigma=5\times 10^{22}$ $s^{-1}$, temperature $T=2\times 10^8$ K and $\mue$ profile $\upmu$1 were used. The highest values of $j_B^2$ within the domain can reach $\sim 50$; for display purposes, we have used the same color for all values above 24. ODAF$=1.14$ for these simulations.}
\label{fig:JSquaredContourOhm1}
\end{figure}

\begin{figure}
\includegraphics[width=\columnwidth]{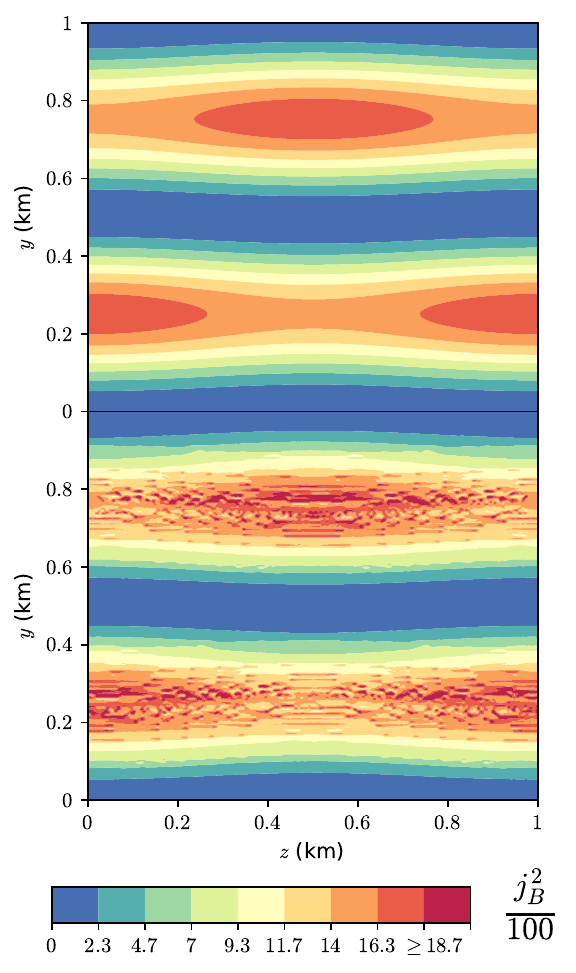}
\caption{Same as Figure~\ref{fig:JSquaredContourOhm1} but with initial field configuration A6. The highest values of $j_B^2/100$ within the domain can reach $\sim 30$; for display purposes, we have used the same color for all values above 18.7. ODAF$=1$ for these simulations, i.e., no enhancement of Ohmic dissipation.}
\label{fig:JSquaredContourOhm2}
\end{figure}

If the magnetic field is too strong so that only a small number of Landau levels are occupied, the enhancement of the Ohmic dissipation is reduced. To show this, in Figures~\ref{fig:JSquaredContourOhm1} and~\ref{fig:JSquaredContourOhm2} we plot the reduced $B=H$ current density squared $j_B^2=|\overline{\bm{\nabla}}\times\overline{\bm{B}}|^2$ throughout the simulation domain at the end of 500 year Ohmic decay-only simulations, with no LQ and LQ simulations starting from identical initial conditions. This gives a measure of the Ohmic dissipation throughout the simulation volume. Figures~\ref{fig:JSquaredContourOhm1} considers initial field configuration A1, while Figure~\ref{fig:JSquaredContourOhm2} considers configuration A6. The results thus differ only in the overall magnitude of the initial configuration, with $B_m$ for the latter being an order of magnitude larger than the former.

The localized regions of small length-scale current density in the lower panels of Figures~\ref{fig:JSquaredContourOhm1} and~\ref{fig:JSquaredContourOhm2}, whose size is resolution-limited, are a consequence of the small-scale field structure generated by the Landau quantization terms. These regions are responsible for the enhanced field dissipation in the LQ simulations compared to the no LQ simulations. They are continually generated throughout the simulation as the field evolves, new dHvA oscillations appear and generate new small length-scale current density. There are clearly fewer small-scale field features generated by the evolution in the lower panel of Figure~\ref{fig:JSquaredContourOhm2} compared to that of Figure~\ref{fig:JSquaredContourOhm1}. This is because of the smaller number of occupied Landau levels, and hence dHvA oscillations, within the range of $B$ and $\mue$ for initial configuration (because of the much stronger magnetic field). This is reflected in the ODAF values for each set of simulations: for Figure~\ref{fig:JSquaredContourOhm1} it is 1.14, while for Figure~\ref{fig:JSquaredContourOhm2} it is $\approx 1$ i.e., no enhancement. This indicates that the enhanced Ohmic dissipation due to dHvA oscillations will only be active for moderately quantizing fields and not for strongly quantizing fields, which have usually been of greater interest for studies of the effects of Landau quantization on neutron stars.

\subsection{Hall--Ohm simulation}
\label{sec:HallOhmSimulations}

We expect the nonlinear Hall term to generate additional small-scale field structure and hence enhance the dissipation of the magnetic field, as has been observed repeatedly in electron MHD/Hall MHD simulations~\citep{Hollerbach2002,Hollerbach2004,Pons2007,Vigano2012,Vigano2013,
Vigano2021,Dehman2023}. Because the Hall timescale is shorter than the Ohmic timescale for the parameter range used in our simulations, and hence the field evolution is faster with the Hall effect turned on, we expect that the Hall effect will enhance the impact of Landau quantization on the field evolution. As the field evolves, the pattern of the dHvA oscillations changes, itself generating additional small-scale field structure, which evolves under the Hall effect with a shorter Hall time than the initial large-scale field.

Turning on the Hall term reduced the maximum time step that we can use for our simulations by 1-2 orders of magnitude compared to those used in the purely Ohmic dissipation simulations. We will thus focus primarily on simulations for $t\leq 50$ years in this Section. This is comparable to the Hall time for length scale $L=0.2$ km for field configuration A1. Hence, all but the lowest wavenumber Fourier components of the field will evolve for longer than a Hall time during these simulations.

\begin{figure}
\center
\includegraphics[width=0.98\linewidth]{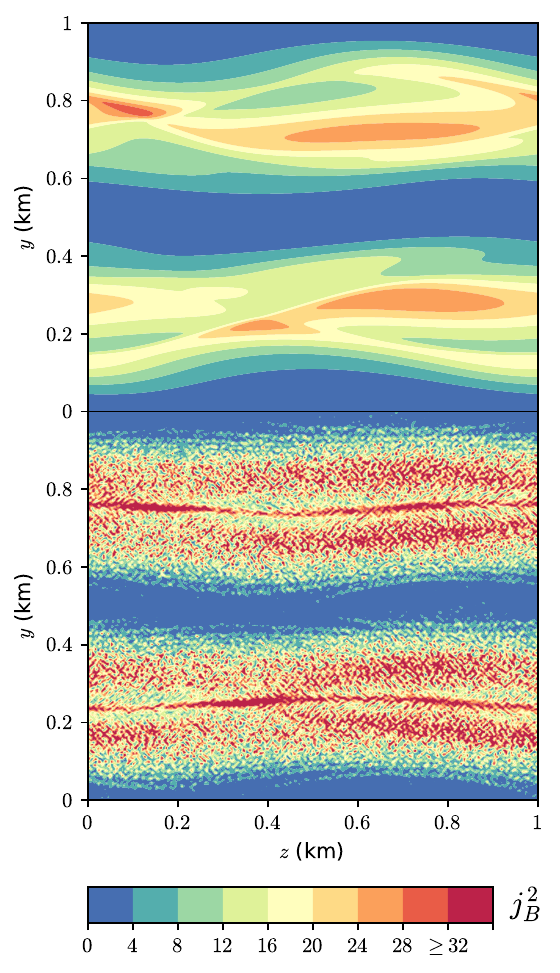}
\caption{Contour plot showing $j_B^2=|\overline{\bm{\nabla}}\times\overline{\bm{B}}|^2$ at the end of 50 yr-long simulations with the Hall term enabled starting from identical initial field configuration A1. Top: Landau quantization effects turned off, $N=128$. Bottom: Landau quantization effects turned on, $N=192$. The small-scale field structure generated by this effect is clearly visible. Fixed conductivity $\sigma=5\times 10^{22}$ $s^{-1}$ and temperature $T=2\times 10^8$ K were used. The highest values of $j_B^2$ within the domain can reach $\sim 100$; for display purposes we have used the same color for all values above 32. ODAF$=1.8$ for these two simulations.} 
\label{fig:JSquaredContourHall}
\end{figure}

Figure~\ref{fig:JSquaredContourHall} shows snapshots of the reduced current density squared $j^2_B$ for two Hall--Ohm simulations with identical initial field configurations, one with no LQ and one with LQ. Compared to Figures~\ref{fig:JSquaredContourOhm1} and~\ref{fig:JSquaredContourOhm2}, the effect of the Hall term on the field evolution is clear, with short length-scale features of the field and hence the current density being generated even in the no LQ case. This is most evident by comparing the top panels of Figures~\ref{fig:JSquaredContourOhm1} and~\ref{fig:JSquaredContourHall}, which use identical initial conditions. Even after ten times as long of a simulation time, the overall field topology has not changed in the top panel of Figure~\ref{fig:JSquaredContourOhm1}, only decayed in strength. The field topology in the top panel of Figure~\ref{fig:JSquaredContourHall}, by contrast, displays strong gradients over short length scales and regions with $j_B^2$ values that are a factor $\sim 1.5$ greater than the maximum value in the top panel of Figure~\ref{fig:JSquaredContourHall}. Comparing the bottom panels of Figures~\ref{fig:JSquaredContourOhm1} and~\ref{fig:JSquaredContourHall}, we see that the small-scale, high $j_B^2$ regions associated with dHvA oscillations of the magnetization and its partial derivatives are smaller in the Hall--Ohm simulation but also cover more of the simulation domain. This follows from the enhanced generation of small-scale field features in the Hall--Ohm simulation compared to the purely Ohmic simulation, and the resulting additional dHvA oscillations that occur, further generating small length-scale field features. The ODAF for these two simulations is 1.8, significantly larger than its value for the otherwise identical Ohmic-only simulation in Figure~\ref{fig:JSquaredContourOhm1}.

\begin{figure}
\center
\includegraphics[width=0.98\linewidth]{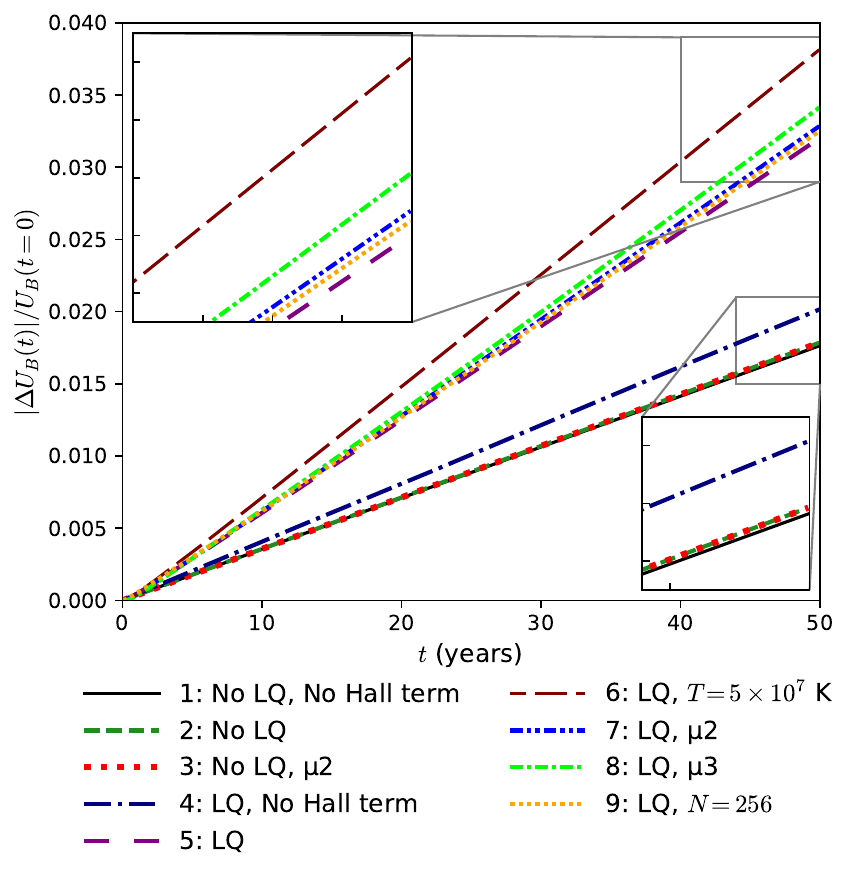}
\caption{Comparison of the fractional magnetic field energy dissipated after $t=50$ years for the A1 field configuration for seven Hall--Ohm simulations and two Ohmic decay-only simulations. Fixed conductivity $\sigma=5\times 10^{22}$ s$^{-1}$ was used for all simulations, as were temperature $T=2\times 10^8$ K, $\mue$ configuration $\upmu$1, and resolution $N=192$ unless otherwise specified.}
\label{fig:DeltaUBHallOhm}
\end{figure}

To quantitatively compare the simulations with and without the Hall term, in 
Figure~\ref{fig:DeltaUBHallOhm} we show the fractional field energy dissipated Eq.~(\ref{eq:FracDissME}) for nine different simulations, all using initial field configuration A1. Among these simulations are two Ohmic decay-only simulations, one no LQ (1) and one LQ (4), for comparison. Notably, we find that including the Hall term does not increase the field energy dissipated for the no LQ simulation (2) compared to an otherwise identical simulation without the Hall term (1), but that the Hall term increases the field energy dissipated for the LQ simulation (5) compared to an otherwise identical simulation without the Hall term (4) by a factor $\approx 1.6$. This is because the Landau quantization terms generate small length-scale field features of significant amplitude immediately, and these features have shorter Hall times than the initial large-scale field. The Hall term thus acts faster for the LQ simulation, and can enhance the field energy dissipated in this simulation when run for times much less than the Hall time of the large-scale initial field. We also see a clear increase in the field energy dissipated as the temperature is decreased by comparing curves 5 and 6, due to the larger amplitude dHvA oscillations at lower temperature. Comparing curves 5 and 9, we observe that increasing the simulation resolution also slight increases the energy dissipated, in contrast to the Ohmic-only simulations where increasing the resolution tended to slightly decrease the dissipated energy. As in Figure~\ref{fig:DeltaUBOhm}, varying the electron chemical potential configuration does not significantly change the field energy dissipated as long as all other variables are held fixed (compare curves 2 to 3 and 5 to 7 and 8), and so we will only consider the $\upmu$1 configuration for the rest of this Section.

\begin{figure}
\includegraphics[width=\columnwidth]{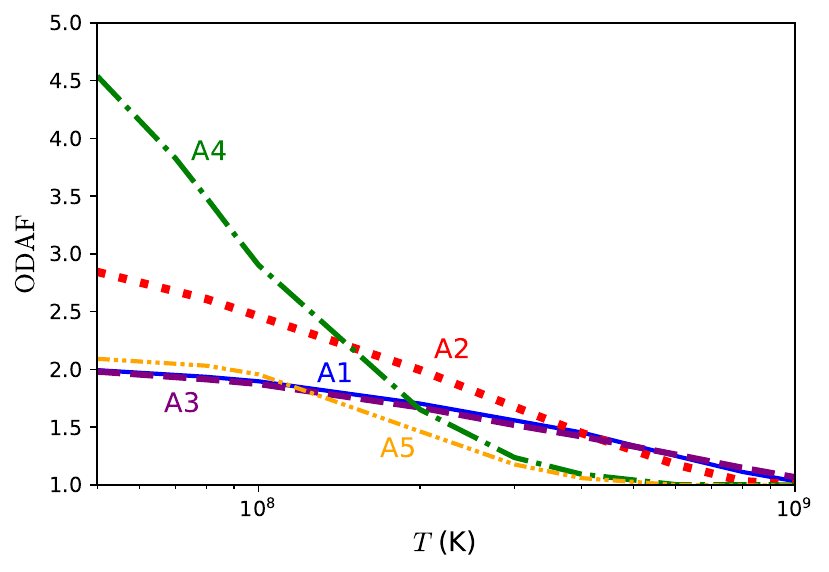}
\caption{Same as Figure~\ref{fig:TempCompOhm} but for Hall--Ohm simulations and with $t=10$ yr simulation times.}
\label{fig:TempCompHall}
\end{figure}

Figure~\ref{fig:TempCompHall} shows the ODAF after 10 yr simulation time as a function of temperature for three different field configurations. Compared to Figure~\ref{fig:TempCompOhm}, we see that the ODAF values can be much larger with the Hall effect, reaching as high as 4.5 for configuration A4 at $T=5\times10^7$ K, compared to a maximum value of around 1.9 at the same temperature for configurations A2 and A4 without the Hall effect. The stronger suppression of the ODAF as the temperature increases for configuration A4 compared to configurations A1 and A2 is reflective of Figure~\ref{fig:MChiM}: at high temperatures, the dHvA oscillations at low $B$ are suppressed much more than they are at high $B$. But because the quantization is weaker at low $B$ and there are more dHvA oscillations to generate the small-scale field structure, as indicated by the wider range in $n_{\rm max}$ for weaker field configurations in Table~\ref{tab:BConfigs}, once the temperature is lowered sufficiently, the ODAF for the weaker field configuration becomes larger than for the stronger field configurations. Configuration A4, which has the lowest average field strength, thus gives the highest ODAF at these low temperatures. The near overlap in the curves for configurations A1 and A3 demonstrates that changing the wavenumber of the initial field (by changing $p_B$) only slightly changes the ODAF value, though it does of course change the absolute value of $\Delta U_B$, with this being about a factor of four larger for configuration A3 than for A1, as expected from $\Delta U_B$ scaling with the wavenumber of the initial field squared. Having larger gradients in $B$ i.e., having larger values of $\beta$ in the initial field configuration, also helps to increase the ODAF at higher temperatures, which is why the ODAF for configuration A5 ($\beta=1$) only starts to increase appreciably below $T\sim 3\times10^8$ K, whereas for configurations A1--A3 ($\beta=10$), the ODAF is already nearly 2 by this temperature. The overall weaker field for $A4$ suppresses the ODAF for this configuration at lower temperatures even if it has a strong field gradient with $\beta=10$. Overall, we find that values of ODAF $\approx 2$ are expected for $1\times10^8 < T < 3\times 10^8$ K for all configurations, which is the temperature range of most interest if Landau quantization-enhanced Joule heating is to help explain the magnetar heating problem.

\begin{figure}
\includegraphics[width=\columnwidth]{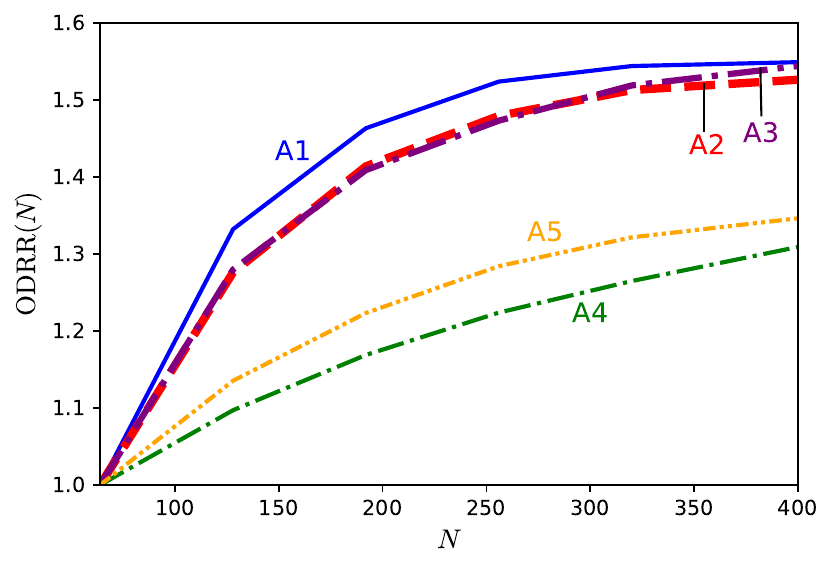}
\caption{Same as Figure~\ref{fig:ResolutionCompOhm} but for Hall--Ohm simulations and with $t=5$ yr simulation times.}
\label{fig:ResolutionCompHall}
\end{figure}

Figure~\ref{fig:ResolutionCompHall} compares the Ohmic dissipation resolution ratio ODRR after 5 yr for different spatial resolutions, for the five different initial field configurations. This ratio appears to converge to $\sim 1.5$ as $N$ is raised above 300 for initial field configurations A1, A2 and A3, with a similar value of ODRR observed for these three configurations, and similar convergence to ODRR$\sim 1.3$ for configurations A4 and A5. This suggests that increasing the resolution may increase the amount of magnetic field energy dissipated by a factor of $\sim1.5$ at most compared to its value at $N=64$, and that resolutions of $N=$a few hundred should give accurate estimates of the enhancement of the Ohmic dissipation due to Landau quantization.

\subsection{Hall cascade spectrum}

As the Landau quantization-induced magnetization terms clearly generate additional small-scale field structure, we now examine how this effect modifies the field energy spectrum. In the absence of Landau quantization effects, this energy spectrum is governed by the transfer of energy to shorter length scales by the nonlinear Hall effect, leading to a Hall cascade~\citep{Goldreich1992,Biskamp1996,Biskamp1999,Ng2003,Cho2004,Cho2009,Wareing2009,
Wareing2010,Gourgouliatos2016,Brandenburg2020,Dehman2023a}. For helical fields, an inverse Hall cascade transferring energy from shorter to longer length scales is also possible due to magnetic helicity conservation~\citep{Cho2011,Brandenburg2020,Gourgouliatos2020}. We do not consider such fields, and thus do not expect to observe the inverse cascade.

We compute the discrete 1D field energy spectrum $E_B^c(k)$ from the spectral coefficients of the magnetic field $\hat{\bm{B}}(\bm{k})$ according to~\citep{Verma2004,Brandenburg2011}
\begin{equation}
E_B^d(k) = \frac{1}{8\pi}\sum_{k_-<|\bm{k}|<k_+}|\hat{\bm{B}}(\bm{k})|^2,
\end{equation}
where $k_{\pm}=k+\delta k/2$ defines the interval around each wavenumber $k$ for which we compute discrete values of $E_B^d(k)$. For $L=L_y=L_z=L_x$ and quasi-2D simulations, $k = 0, 2\pi/L, 4\pi/L, ..., 2\pi\lfloor N/\sqrt{2}\rfloor/L$ \footnote{\texttt{Dedalus} uses cosine and sine basis functions to represent $\bm{A}$, and so there are $N/2$ distinct values of $k$ used in each spatial direction with resolution $N$. This is then multiplied by $\sqrt{2}$ for the two spatial directions since $k_{\rm max}=\sqrt{k_{y,{\rm max}}^2+k_{z,{\rm max}}^2}$ in the isotropic turbulence approximation.} and $\delta k = 2\pi/L$. The energy spectrum is related to the volume-averaged magnetic energy density by
\begin{equation}
\langle u_B\rangle=\frac{\langle B^2\rangle}{8\pi}=\sum_{k=0}^{2\pi\lfloor N/\sqrt{2}\rfloor/L}E_B^d(k),
\end{equation}
and the usual (continuous) 1D field energy spectrum $E_B(k)$ is defined such that
\begin{equation}
\int E_B(k){\rm d}k = \langle u_B\rangle \rightarrow E_B(k) = \frac{1}{k_0}E_B^d(k),
\end{equation}
where $k_0 = 2\pi/L$.

\begin{figure}
\includegraphics[width=\columnwidth]{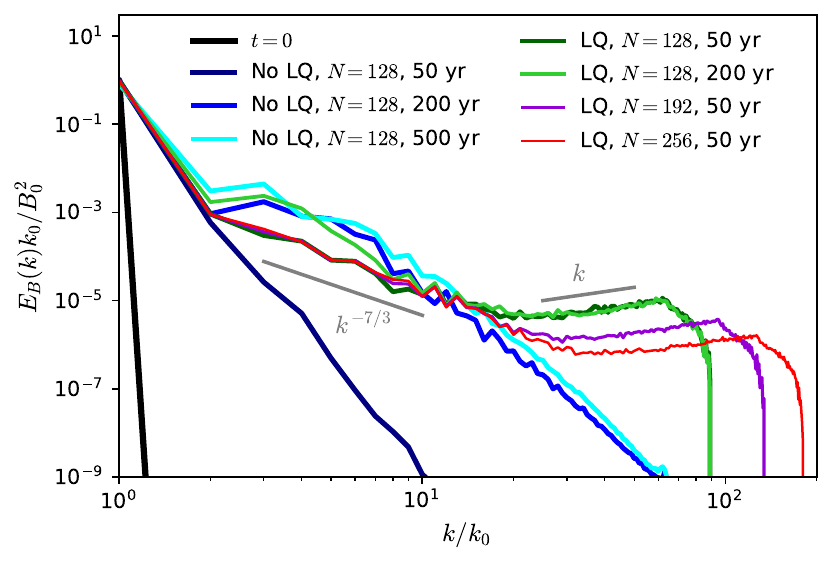}
\caption{Field energy spectrum $E_B(k)$ for identical initial coherent field configuration A1 for different resolutions $N$, where $k_0=2\pi/L$. The spectrum for one no LQ and three LQ simulations with different resolutions are displayed at different times. Fixed conductivity $\sigma=5\times10^{22}$ s$^{-1}$, temperature $T=2\times10^8$ K and $\mue$ profile $\upmu$1 were used for all simulations. Straight lines show the expected scaling of the spectra for the Hall cascade $\propto k^{-7/3}$ and $\propto k$ at large $k$ for the no LQ simulations.} 
\label{fig:E_BSpectrumLargeScaleB}
\end{figure}

We first consider the field energy spectrum for simulations involving an initial large-scale (coherent) field. Figure~\ref{fig:E_BSpectrumLargeScaleB} shows the field energy spectrum using initial field configuration A1 for different times for a No LQ and LQ simulations at different resolutions. By comparing the No LQ to the LQ simulations after $t=50$ yr, it is clear that in the LQ simulations, small length-scale field structure is very quickly generated. This additional small-scale field structure leads to enhanced field dissipation, with an ODAF of 1.83 computed by comparing the No LQ simulation to the $N=256$ LQ simulation after 50 yr. This takes the form of a distinct upward-sloping plateau $\propto k$ as the resolution limit is approached. This plateau is only seen when the Landau quantization effects are included, and is generated by the joint action of the Hall term and dHvA oscillations on the field evolution. Its peak occurs at $k/k_0=N/2$, the largest wavenumber in a single direction represented in the simulations. The spectrum is sharply cut off at $k/k_0=\sqrt{2}N$, the largest resolved wavenumber. The expected scaling for a turbulent Hall cascade in electron MHD, $E_B(k)\propto k^{-7/3}$ ~\citep{Biskamp1996,Biskamp1999,Ng2003,Cho2004,Cho2009,Brandenburg2020}, is shown for comparison. This scaling is only satisfied for a range $2<k/k_0\lesssim k_b$ for the LQ simulations, where $k_b$ denotes the break in the spectrum between the downward slope and the plateau. The Hall cascade develops much faster for the no LQ simulation because of the immediate generation of large $k$ field components, whereas it develops slowly for the no LQ simulations since the Hall time for $k=k_0$ for the simulation is of the order of 80 yr. For the no LQ simulation, $E_B(k)\propto k^{-7/3}$ loosely holds for $2<k/k_0\lesssim 15$ for $t\geq 200$ yr, with Ohmic dissipation giving the spectrum a steeper slope at larger $k$ where dissipation is faster.

To properly compare our simulations to the expected Hall cascade scaling and to determine how this scaling is modified by the Landau quantization-induced magnetization, we require a turbulent initial field. To generate this, we use as our initial vector potential a quasi-flat spectrum over the range $5\leq k/k_0 \leq 10$, assigning to all coefficients in the 2D sine-cosine Fourier basis the value
\begin{align}
A_i(\bm{k}) = B_m\left(a_i-\frac{k_i(k_ja^j)}{k^2}\right)S(k),
\label{eq:RandomA}
\end{align}
where $a_i=(\delta_i^x+\delta_i^y+\delta_i^z)/\sqrt{3}$, $k_x=0$ for our quasi-2D simulations and $B_m=10^{15}$ G. The initial spectral shape $S(k)$ is
\begin{equation}
S(k) = \frac{1}{k}\left[\theta\left(10-\frac{k}{k_0}\right)-\theta\left(5-\frac{k}{k_0}\right)\right],
\end{equation}
where $\theta(k)$ is the Heaviside step function. Using Eq.~(\ref{eq:HallTimescale}) with $L\rightarrow 1/k$, $n_{\rm e}=1.9\times10^{-4}$ fm and $B=5\times 10^{15}$ G, the Hall time for the largest initial wavenumber in the initial condition $k=10k_0$ is thus 0.61 years.

\begin{figure}
\includegraphics[width=\columnwidth]{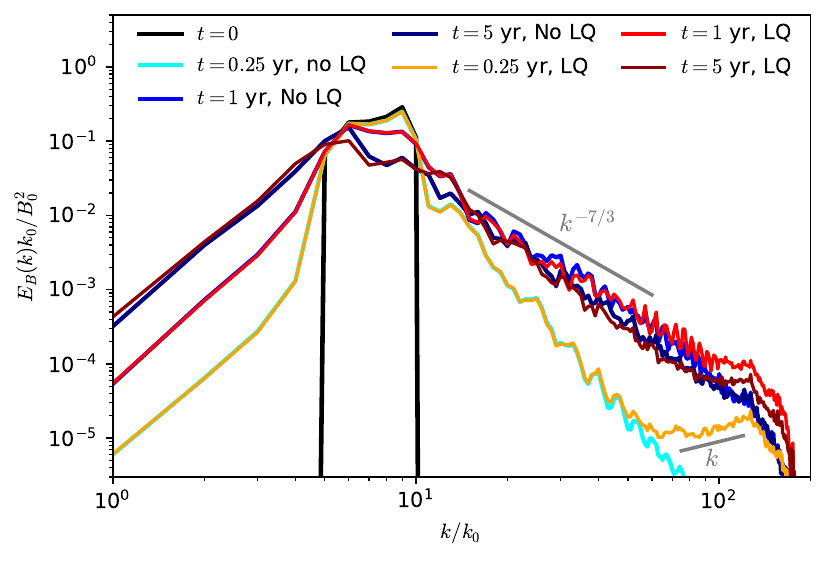}
\caption{Field energy spectrum $E_B(k)$ for quasi-flat initial field configuration for one no LQ and one LQ simulation. The $t=0$ spectrum is shown alongside the spectra at three different times $t=0.25,1$ and $5$ years for both simulations. Identical resolutions $N=256$, fixed conductivity $\sigma=5\times10^{22}$ s$^{-1}$ and $\mue$ profile $\upmu$1 were used for both simulations, and temperature $T=2\times10^8$ K was used for the LQ simulation. Straight lines show the scaling of the Hall cascade $\propto k^{-7/3}$ and $\propto k$ at large $k$ for the no LQ simulation at early time.} 
\label{fig:E_BSpectrumRandomB}
\end{figure}

Figure~\ref{fig:E_BSpectrumRandomB} shows the magnetic energy spectrum for different times with and without Landau quantization effects, with Eq.~(\ref{eq:RandomA}) used as the initial vector potential and with identical resolution. The expected $k^{-7/3}$ spectrum is reproduced within 1 yr for both simulations, and contrary to the initial large-scale field, the Hall effect also moves field energy to longer length scales. For longer simulation times, the dissipation of the field becomes sufficiently important to steepen the scaling at larger wavenumbers, as one can see beginning by $t=5$ yr: at this time, over 8 Hall times have passed for $k=10k_0$, with a significant fraction of the field energy being transferred to large wavenumbers by the Hall cascade to be Ohmically dissipated on shorter timescales than at $k=10k_0$. The spectrum for the simulations with LQ also features a plateau $\propto k$ as the resolution limit is approached similarly to that observed in Figure~\ref{fig:E_BSpectrumLargeScaleB}. This upward slope becomes harder to distinguish over time as the Hall cascade is fully established and extends to higher $k$, washing out the effect of the dHvA oscillations of the magnetization. These results suggest that for turbulent initial conditions, the additional small-scale field structure generated by Landau quantization through dHvA oscillations does not significantly enhance Ohmic dissipation of the field. This is supported by similarity of the amount of magnetic field energy dissipated by each simulation: the ODAF at $t=5$ yr for them is only 1.08, with both simulations resulting in approximately $1/3$ of the initial field energy being dissipated.

\begin{figure}
\includegraphics[width=\columnwidth]{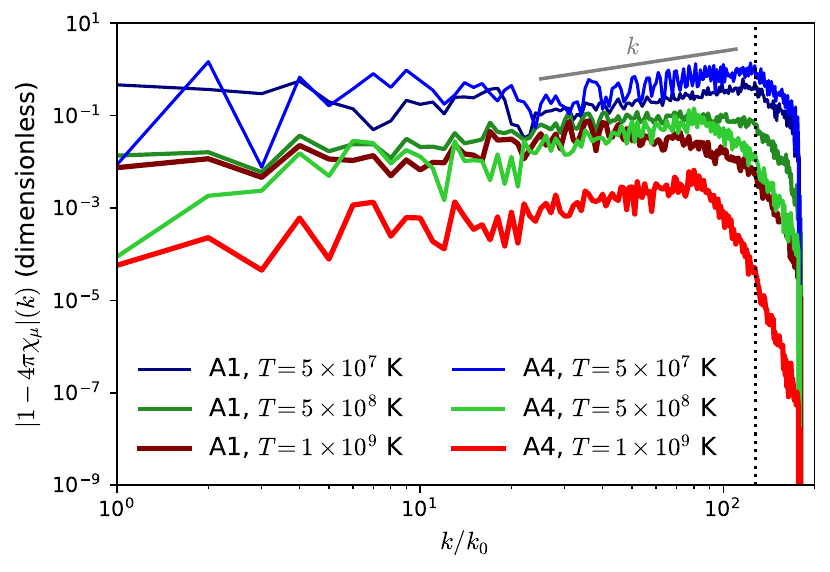}
\caption{Fourier spectrum for $|1-4\pi\chi_{\mu}|$ for initial field configurations A1 and A4 at resolution $N=256$, with $\mue$ configuration $\upmu1$ and three different temperatures. The resolution cutoff at $k/k_0=N/2$ is given by a dotted line, showing that $|1-4\pi\chi_{\mu}|(k)$ peaks before this value for the $T=5\times 10^8$ and $1\times 10^9$ K temperature curves. The spectra generally scale as $k$ below their peak.}
\label{fig:ChiMuSpectrum}
\end{figure}

The spectra for both the coherent and turbulent initial field configurations display a similar $\sim k$ plateau for large wavenumbers independent of the resolution before dropping off at the resolution limit. The standard Hall cascade spectrum is
\begin{equation}
E_B(k) \sim \frac{B_k^2}{k} = C_H \epsilon^{2/3}k^{-7/3},
\end{equation}
where $B_k$ is the magnetic field component with wavenumber $k$, $C_H$ is a constant of order unity, and $\epsilon$ is the spectral energy cascade rate. The usual assumption of Kolmogorov-type turbulence analysis is that $\epsilon$ is wavenumber-independent in the inertial subrange: clearly, for our system, this assumption no longer holds for sufficiently large wavenumbers above the break in the spectrum at $k_b$. Above $k_b$, to obtain $E_B(k)\sim B(k)^2/k\sim k$ requires that $B_k\sim k$: why would this be? We know that the main difference between the No LQ and LQ simulations is the inclusion of the term depending on $\chi_{\mu}$ in Eq.~(\ref{eq:JMagnetization}). In Figure~\ref{fig:ChiMuSpectrum} we show the spectrum of $|1-4\pi\chi_{\mu}|(k)$ for two different large-scale initial field configurations A1 and A4 at different temperatures. These spectra show $\propto k$ scaling at low temperatures until the resolution-limited cutoff. As the temperature increases, the cutoff becomes lower than the resolution limit, but $|1-4\pi\chi_{\mu}|(k)$ still scales linearly in $k$ for $k$ below the peak for the temperatures and field configurations we examined. The temperature suppression of $|1-4\pi\chi_{\mu}|(k)$ is more significant for A4 than A1 because the field strength is factor of $5$ weaker for the former. These results suggest that if $B_k\propto |1-4\pi\chi_{\mu}|(k)\propto k$ above $k_b$, which follows from Eq.~(\ref{eq:MagneticInduction}) and~(\ref{eq:JMagnetization}), then $E_B(k)\sim k$ here.

What is $k_b$? It occurs when the Hall cascade from small $k$ intersects the $\propto k$ plateau due to dHvA oscillations at large $k$. This is a time-dependent quantity, as Figure~\ref{fig:E_BSpectrumRandomB} clearly demonstrates: only in the $t=0.25$ yr, LQ spectrum is this plateau clearly visible, since for latter times the Hall cascade is sufficiently developed to overwhelm it. Figure~\ref{fig:E_BSpectrumLargeScaleB} shows that $k_b$ occurs at larger values for higher resolutions because the amplitude of the plateau decreases as resolution increases, so it intersects with the Hall cascade at higher $k$. The amplitude of the plateau decreases with resolution because the dHvA oscillations can be more accurately represented using a larger spectral basis including higher wavenumbers, so as $N$ increases, the plateau in $E_B(k)$ becomes lower in amplitude but broader in $k$. At a sufficiently high resolution, which will be a function of the temperature as Figure~\ref{fig:ChiMuSpectrum} suggests, the $\sim k$ plateau should be fully resolved and terminate without being forced to at the highest resolved $k$.

\section{Discussion and Conclusion}
\label{sec:Conclusion}

The magnetic fields attained in magnetar magnetospheres and crusts are sufficiently strong that electron quantization into a single or moderate number of Landau levels will occur. Landau quantization has been shown in previous studies to have dramatic effects on electron transport, neutrino emissivity and thermodynamic properties such as differential magnetic susceptibility and the specific heat capacity. In this study, for the first time we include the effects of Landau quantization on the magnetization and its partial derivatives in numerical simulations of magnetic field evolution in conditions typical of the crust of strongly magnetized neutron stars.

Our primary aim was to study the viability of a proposed mechanism to enhance the Ohmic dissipation of crustal magnetic fields by generating small-scale field structures associated with the dHvA oscillations of the magnetization and its partial derivatives $\chi_{\mu}$ and $\mathcal{M}_{\mu}$ as defined in Eq.~(\ref{eq:MagnetizationPartialDerivatives}). To this end, we performed a series of numerical experiments in which we simulated magnetic field evolution in a periodic domain assuming conditions appropriate to the inner crust ($\mu_e\approx35$ MeV) with various models for the initial magnetic field. We have shown that this mechanism does indeed increase the Ohmic dissipation of the magnetic field by a factor up to $\sim 5$ compared to the case without Landau quantization and dHvA oscillations. It does so in three ways: (1) by decreasing the effective electrical conductivity perpendicular to the magnetic field (see Eq.~(\ref{eq:JMagnetizationSplit})); (2) by increasing the effective Hall diffusivity, so the Hall cascade (which converts long length-scale field to shorter length-scale, more rapidly dissipated field) is faster; and (3) by directly generating short-wavelength field features, which undergo Ohmic decay and Hall drift more rapidly than the initial large-scale field components (as demonstrated in Figure~\ref{fig:DeltaUBHallOhm}).

The exact enhancement factor of the Ohmic decay compared to the case ignoring Landau quantization and dHvA oscillations depends on the initial field configuration and the temperature. Higher temperatures mitigate the effect because they reduce the amplitude of the dHvA oscillations. We find that field configurations that are moderately quantizing--in which electrons are quantized into hundreds of Landau levels across the simulation domain--give the greatest enhancement to the Ohmic decay, as long as temperatures are sufficiently low that the dHvA oscillations are not thermally suppressed. These configurations also show the largest growth in their enhancement factor as the temperature decreases. They have the largest-amplitude dHvA oscillations at low temperatures, and the oscillations are densely spaced within the simulation domain. In contrast, strongly quantizing fields ($n_{\rm max}\lesssim 5$) provide small enhancement factors, since the amplitude of the dHvA oscillations of $M$, $\chi_{\mu}$ and $\mathcal{M}_{\mu}$ are low and occur sparsely across the simulation domain. 

\citet{Beloborodov2016} argued that Ohmic dissipation of crustal fields could not resolve the magnetar heating problem unless the magnetic fields had extreme variations of order $10^{16}$ G over $\sim 100$ m length scales, since only such variations could provide sufficient heating to account for inferred magnetar surface temperatures. The mechanism proposed in this paper generates field fluctuations that are only $\sim10^{14}$--$10^{15}$ G in magnitude, but over length scales much smaller than $100$ m. The Joule heating rate is estimated as
\begin{equation}
\dot{q}\sim 6\times10^{18}\left(\frac{|\bm{\nabla}\times\bm{B}|}{10^{11}{\ \rm G\ cm}^{-1}}\right)^2\left(\frac{10^{22}{\rm s}^{-1}}{\sigma}\right){\ \rm erg\ s}^{-1}{\rm cm}^{-3},
\end{equation}
with $\dot{q}\gtrsim 3\times10^{19}$ erg s$^{-1}$ cm$^{-3}$ required to explain inferred magnetar surface temperatures. We assume that the local fluctuations in the effective conductivity due to the dHvA oscillations of $M$ and $\chi_{\mu}$ are included in the value of $\sigma$. Based on Figure~\ref{fig:JSquaredContourHall}, values of $|\bm{\nabla}\times\bm{B}|\sim 6B_0/L_0=6\times10^{11}$ G cm$^{-1}$ are typical for the dHvA oscillation-induced maxima in $j_B^2$, which would have local heating rate $\dot{q}\sim 2\times10^{20}$ erg s$^{-1}$ cm$^{-3}$, exceeding the required heating rate by an order of magnitude. This extreme heating only occurs over a fraction $\lesssim$1/5 of the simulation volume, reducing the heating rate to $\dot{q}\sim 4\times10^{19}$ erg s$^{-1}$ cm$^{-3}$. This is marginally sufficient to explain magnetar surface temperatures, and will be increased by weaker heating occurring over much of the simulation volume. Notably, the small-scale field features that result in the strongest localized Joule heating are continually regenerated as long as the magnetic field is sufficiently strong for a given temperature for them to not be thermally suppressed, so this enhanced heating rate could be expected to persist for long periods of the magnetar's lifetime. The decay is further enhanced by the Hall effect, which acts on shorter times when dHvA oscillations can generate small wavelength field features that have shorter Hall times than the large-scale field. These arguments combine to suggest the viability of Landau quantization-enhanced Ohmic decay as an explanation for the hotter inferred temperatures of magnetars compared to other classes of neutron stars.

Our simulations were unable to resolve all of the dHvA oscillations of $M$, $\chi_{\mu}$ and $\mathcal{M}_{\mu}$, and our estimate in Eq.~(\ref{eq:dHvALengthScale}) suggests that resolving all oscillations in the interesting moderately quantizing regime is not computationally feasible. However, we were able to show that the resolution-dependence of the Ohmic dissipation enhancement factor tends to saturate at realistic resolutions. Thus, simulations can capture enough of the field dynamics associated with dHvA oscillations that these effects could be included in realistic simulations of neutron star crusts, as opposed to the periodic box domain used in this paper. One key factor that we neglected in this work is temperature evolution, instead using fixed temperature. We found that lower temperatures increase the dHvA-enhanced Ohmic dissipation the most: in realistic simulations, electrical conductivity would also increases with decreasing temperature, somewhat offsetting this enhancement. The increased heating due to dHvA-enhanced Ohmic dissipation would also increase the temperature, thus reducing the enhanced Ohmic dissipation and resulting heating. Additionally, the SdH oscillations and anisotropy of the resistivity, which we ignored in this paper, could also potentially increase the conductivity/lower Ohmic dissipation in parts of the simulation domain, and could also decrease the conductivity/enhance Ohmic dissipation in other parts of the simulation domain. Without running the simulations, we cannot make a firm prediction about whether the dHvA-enhancement of the Ohmic dissipation could be offset by the SdH oscillations or anisotropy of the electrical conductivity. Since the dHvA oscillations will still provide an enhancement to the Hall diffuvisity and help to generate additional small-scale field structure regardless of the SdH oscillations of the conductivity, it is unlikely that the SdH oscillations will entirely suppress the enhanced Ohmic dissipation of the field we observed here. Regardless, it is clear that fully understanding the effect that Landau quantization will have on neutron stars requires including these effects in joint magneto-thermal evolution simulations.

Other extensions of this work should consider the effect of magnetic field evolution including dHvA oscillations of the magnetization and its partial derivatives on the mechanical properties of neutron star crusts. The assumption of magnetohydrostatic balance for the crust could be relaxed to give a more realistic picture of the dynamics when the mechanical failure of the crust occurs. In magnetar crusts, magnetic stresses may be sufficiently strong to cause the crust to undergo plastic failure, in which case the criterion for electron MHD is not strictly satisfied. There has been much recent progress in the study of plastic failure in neutron star crusts~\citep{Li2016,Lander2019,Kojima2020,Gourgouliatos2022}. However, plastic flow does not appear to completely suppress the Hall effect~\citep{Gourgouliatos2021}, so neglecting it in our field evolution simulations is a reasonable first step.
\\
\\
P.B.R. was supported by the INT's U.S. Department of Energy grant No. DE-FG02-00ER41132, and the Simons Foundation through a SCEECS postdoctoral fellowship (grant No. PG013106-02). He would like to thank the \texttt{Dedalus} developers for advice on the numerics and for making their code publicly available. He would also like to acknowledge B. P. Brown (bpbrown) and E. Kaufman (ekaufman5) for making Dedalus scripts publicly available on Github, which were referenced in writing the scripts for this project, and V. Skoutnev for advice on running \texttt{Dedalus} on Columbia's HPC resources. The authors would also like to thank A. Brandenburg, Y. Levin, and the anonymous referee for helpful discussion and comments. All plots were made using the Python package \texttt{matplotlib}~\citep{Matplotlib}. The scripts used in this paper are available at https://github.com/PRauNS/ElectronMHDpublic.

\vspace{5mm}


\appendix

\section{Temperature-dependent expressions for required thermodynamic functions}
\label{app:Thermodynamics}

Our magneto-thermal simulations require computing temperature and magnetic field-dependent thermodynamic functions $M$, $\chi_{\mu}$ and $\mathcal{M}_{\mu}$. In the nonmagnetized case, the Sommerfeld expansion is generally sufficient to compute the finite temperature corrections to the $T=0$ Fermi gas thermodynamic functions. This approximation is only partially applicable to the $B\neq0$ case as discussed in~\citet{Rau2023}, so to avoid computationally expensive Fermi--Dirac integrals, we developed a set of different approximations to compute the required thermodynamic functions at finite temperature. In this Appendix we work in Gaussian units and set $\hbar=c=k_{\rm{B}}=1$.

The most convenient thermodynamic potential for the purposes of our macroscopic physical description is the grand potential density $\Omega$; here, we are only concerned with $\Omega$ for the electrons, $\Omega_{\rm e}=\Omega_{\rm e}(\mue,T,B)$. The vacuum magnetic field energy density $B^2/(8\pi)$ must be included in the full expression for $\Omega$, but is not included in $\Omega_{\rm e}$. At nonzero $T$, including spin coupling to the magnetic field and assuming the electron $g$-factor is exactly 2, $\Omega_{\rm e}$ is given by~\footnote{We ignore the effects of the crystal lattice and electron band structure on the Landau quantization of electrons, and ignore the possibility of exciting positrons since the relevant temperatures for our simulations are far too small to excite them in any significant amount.}
\begin{align}
\Omega_{\rm e}(\mue,T,B)=-\frac{eB}{4\pi^2}\sum_{n=0}^{\infty}g_n{}&\int_{-\infty}^{\infty}{\rm d}p\frac{p^2}{\sqrt{p^2+m_n^2}}f_F(p,\mue,B,n,T),
\label{eq:GrandPotentialDensity}
\end{align}
where $g_n=2-\delta_{n,0}$ is a degeneracy factor, $m_n^2=m_{\rm e}^2+2eBn$, $m_{\rm e}$ is the electron mass and $f_F$ is the Fermi--Dirac distribution
\begin{equation}
f_F(p,\mue,B,n,T)=\frac{1}{e^{\beta(\sqrt{p^2+m_n^2}-\mue)}+1},
\end{equation}
where $\beta=1/T$. At $T=0$ this reduces to
\begin{equation}
\Omega_{\rm e}(\mue,T=0,B)=-\frac{eB}{4\pi^2}\frac{}{}\sum_{n=0}^{n_{\text{max}}}g_n\left[\mue E_{Fn}-m_n^2\ln\left(\frac{\mue+E_{Fn}}{m_n}\right)\right],
\label{eq:GrandPotentialDensityT=0}
\end{equation}
where $E_{Fn}=\sqrt{\mue^2-m_{\rm e}^2-2eBn}$ and $n_{\text{max}}=\lfloor(\mue^2-m_{\rm e}^2)/(2eB)\rfloor$ is the highest occupied Landau level at $T=0$ respectively. The sum over Landau levels $n$ can be approximated to good accuracy using the Euler--Maclaurin summation formula (e.g.,~\citet{Abramowitz1972}). The required partial derivatives of $\Omega_{\rm e}$ for the fixed temperature magnetic field evolution simulations in this paper are
\begin{subequations}
\begin{align}
M ={}& -\left.\frac{\partial \Omega_{\rm e}}{\partial B}\right|_{\mue,T},
\label{eq:Magnetization}
\\
\chi_{\mu} ={}& -\left.\frac{\partial^2\Omega_{\rm e}}{\partial B^2}\right|_{\mue,T},
\label{eq:chi_mu}
\\
\mathcal{M}_{\mu} ={}& -\left.\frac{\partial^2\Omega_{\rm e}}{\partial B\partial\mue}\right|_{T}.
\label{eq:M_mu}
\end{align}
\end{subequations}
To include finite $T$ in these expressions, we can use a set of two approximations: one for $T\leq eB/(2\pi^2\mu_{\rm e})$ and one for $T>eB/(2\pi^2\mu_{\rm e})$.

\subsection{Low-temperature Approximations}

For low temperatures $T\leq eB/(2\pi^2\mu_{\rm e})$, we can note that only a few Landau levels above $n=n_{\text{max}}$ will be occupied, and the Landau levels below $n=n_{\text{max}}$ will be essentially unaffected. We can thus use the $T=0$ result Eq.~(\ref{eq:GrandPotentialDensityT=0}) for $n<n_{\text{max}}$, and then include finite $T$ corrections only for $n=n_{\text{max}},n_{\text{max}}+1$ to good approximation.

We derive in detail only the expression for $\partial^2\Omega_{\rm e}/\partial B\partial\mu_e$ and merely list the results for the low-temperature approximations for the other required thermodynamic partial derivatives. For $n<n_{\text{max}}$ and defining $n'\equiv n_{\text{max}}-1$ we have
\begin{align}
\left(\frac{\partial^2\Omega_{\rm e}}{\partial B\partial\mu_{\rm e}}\right)_{n<n_{\text{max}}} ={}& -\frac{e}{2\pi^2}\sum_{n=0}^{n'}g_n\frac{E_{Fn}-eBn}{E_{Fn}}
\nonumber
\\
= {}& -\frac{ep_F}{2\pi^2} - \frac{e}{\pi^2}\left(n'E_{Fn'}-E_{F1}\right) 
 - \frac{e}{2\pi^2}\left( \frac{E_{Fn'}-eBn'}{E_{Fn'}} + \frac{E_{F1}-eB}{E_{F1}} \right) - \mathcal{B}_2\frac{e^2B}{2\pi^2}\left( \frac{2E_{F1}+eB}{E_{F1}^3} - \frac{2E_{Fn'}+eBn'}{E_{Fn'}^3}\right)
\nonumber
\\
{}& - \mathcal{B}_4\frac{e^4B^3}{8\pi^2}\left( \frac{4E_{F1}+5eB}{E_{F1}^7} - \frac{4E_{Fn'}+5eBn'}{E_{Fn'}^7}\right) + ...
\end{align}
where $p_F=\sqrt{\mu_{\rm e}^2-m_{\rm e}^2}$ and we used the Euler--Maclaurin summation formula, explicitly excluding the $n=0$ term due to its different degeneracy factor. $\mathcal{B}_2=1/6$ and $\mathcal{B}_4=-1/30$ are Bernoulli numbers. For the $n=n_{\text{max}}$ and $n_{\text{max}}+1$ terms, we start from the finite temperature expression
\begin{align}
\left(\frac{\partial^2\Omega_{\rm e}}{\partial B\partial\mu_{\rm e}}\right)_{n\geq n_{\text{max}}} ={}& \frac{e\beta}{2\pi^2}\sum_{n_{\text{max}}}^{n_{\text{max}}+1}g_n\int_{m_n}^{\infty}{\rm d}E\frac{E^2-m_n^2-eBn}{\sqrt{E^2-m_n^2}}\frac{{\rm e}^{\beta(E-\mu_{\rm e})}}{\left({\rm e}^{\beta(E-\mu_{\rm e})}+1\right)^2}
\nonumber
\\
={}& -\frac{e}{2\pi^2}\sum_{n_{\text{max}}}^{n_{\text{max}}+1}g_n\int_{m_n}^{\infty}{\rm d}E\frac{E}{\sqrt{E^2-m_n^2}}\left(1+\frac{eBn}{E^2-m_n^2}\right)\frac{1}{{\rm e}^{\beta(E-\mu_{\rm e})}+1},
\label{eq:dOmegadBdmuPrelim}
\end{align}
where $m_n=\sqrt{m_e^2+2eBn}$. The second form of Eq.~(\ref{eq:dOmegadBdmuPrelim}) is obtained upon an integration by parts after changing the integration variable to $p=\sqrt{E^2-m_n^2}$ and then changing the integration variable back to $E$. Upon a change of variable $x=E/m_n$ and defining $a\equiv\beta m_n$ and $b\equiv\beta\mue$, we obtain
\begin{align}
\left(\frac{\partial^2\Omega_{\rm e}}{\partial B\partial\mu_{\rm e}}\right)_{n\geq n_{\text{max}}} ={}& -\frac{e}{2\pi^2}\sum_{n_{\text{max}}}^{n_{\text{max}}+1}g_nm_n\left[\mathcal{I}_1(a,b)+\frac{eBn}{m_n^2}\mathcal{I}_2(a,b)\right],
\label{eq:dOmegadBdmu}
\\
\mathcal{I}_1(a,b)\equiv{}& \int_1^{\infty}{\rm d}x\frac{x}{\sqrt{x^2-1}}\frac{1}{{\rm e}^{ax-b}+1},
\label{eq:I1}
\\
\mathcal{I}_2(a,b)\equiv{}& \int_1^{\infty}{\rm d}x\frac{x}{(x^2-1)^{3/2}}\frac{1}{{\rm e}^{ax-b}+1}.
\label{eq:I2}
\end{align}
For small values of $a$ and $b$, we create 2D interpolating functions to compute $\mathcal{I}_1$ and $\mathcal{I}_2$. Since there is a singularity within $\mathcal{I}_2$ in this form, to generate its interpolating function, we use the form
\begin{equation}
\mathcal{I}_2(a,b)=-a\int_1^{\infty}\frac{1}{\sqrt{x^2-1}}\frac{{\rm e}^{ax-b}}{({\rm e}^{ax-b}+1)^2},
\end{equation}
obtained by an integration by parts. We use the interpolating functions for $\mathcal{I}_1$ and $\mathcal{I}_2$ for $\sqrt{a^2+b^2}<50$.

For other values of $a$ and $b$, it is much more efficient and accurate to determine approximate forms of the functions. First, for $b\gg a$, $I_1$ becomes
\begin{align}
\mathcal{I}_1^{b\gg a} ={}& \int_1^{b/a}{\rm d}x\frac{x}{\sqrt{x^2-1}}-\int_1^{b/a}{\rm d}x\frac{x}{\sqrt{x^2-1}}\frac{1}{{\rm e}^{b-ax}+1}+\int_{b/a}^{\infty}{\rm d}x\frac{x}{\sqrt{x^2-1}}\frac{1}{{\rm e}^{ax-b}+1},
\nonumber
\\
\approx{}& \sqrt{\frac{b^2}{a^2}-1}+\frac{1}{a^2}\int_0^{\infty}{\rm d}s\left(\frac{b+s}{\sqrt{(b+s)^2/a^2-1}}-\frac{b-s}{\sqrt{(b-s)^2/a^2-1}}\right)\frac{1}{{\rm e}^s+1}
\nonumber
\\
\approx & \sqrt{\frac{b^2}{a^2}-1}-\frac{1}{a}\int_0^{\infty}{\rm d}s\left(\frac{2a^2}{(b^2-a^2)^{3/2}}s+\frac{a^2(4b^2+a^2)}{(b^2-a^2)^{7/2}}s^3\right)\frac{1}{{\rm e}^s+1}
= \sqrt{\frac{b^2}{a^2}-1}-\frac{\pi^2}{6}\frac{a}{(b^2-a^2)^{3/2}}-\frac{7\pi^4}{120}\frac{a(4b^2+a^2)}{(b^2-a^2)^{7/2}},
\end{align}
where we set $s=b-ax$ and $s=ax-b$ in the second and third terms going from the first to second steps and approximated $b-a\rightarrow\infty$ in the upper bound of integration in the second term. In the third line, we Taylor-expanded the integrand for $s/b\ll 1$. Likewise for $I_2$, using the form of Eq.~(\ref{eq:I2}) we have
\begin{equation}
\mathcal{I}_2^{b\gg a}\approx -\frac{a}{\sqrt{b^2-a^2}}-\frac{\pi^2}{6}\frac{a(2b^2+a^2)}{(b^2-a^2)^{5/2}}-\frac{7\pi^4}{120}\frac{a(8b^4+24a^2b^2+3a^4)}{(b^2-a^2)^{9/2}}.
\end{equation}
Hence, in this limit Eq.~(\ref{eq:dOmegadBdmu}) becomes
\begin{align}
\left(\frac{\partial^2\Omega_{\rm e}}{\partial B\partial\mu_{\rm e}}\right)^{b \gg a}_{n\geq n_{\text{max}}} = -\frac{e}{2\pi^2}\sum_{n=n_{\text{max}}}^{n_{\text{max}}+1}g_n\Bigg[{}&\frac{\mue^2-m_{\rm e}^2-3eBn}{\sqrt{\mue^2-m_n^2}}
-\frac{\pi^2}{6}\frac{\mue^2(m_{\rm e}^2+4eBn)-m_n^2(m_{\rm e}^2+eBn)}{(\mue^2-m_n^2)^{5/2}}T^2
\nonumber
\\
{}& -\frac{7\pi^4}{120}\frac{\mue^4(4m_n^2+8eBn)+3\mue^2m_n^2(8eBn-m_n^2)+m_n^4(3eBn-m_n^2)}{(\mue^2-m_n^2)^{9/2}}T^4\Bigg].
\end{align}
These approximations are equivalent to the zero-temperature terms plus Sommerfeld expansion corrections. We use them for $b-a>50$.

For $a>b$, we use the formula for the sum of a geometric series to obtain for $I_1$
\begin{equation}
\mathcal{I}_1^{a>b} = \int_1^{\infty}{\rm d}x\frac{x}{\sqrt{x^2-1}}\sum_{q=1}^{\infty}(-1)^{q-1}{\rm e}^{-q(ax-b)}=\sum_{q=1}^{\infty}(-1)^{q-1}{\rm e}^{qb}K_1(aq)
\end{equation}
where $K_1(x)$ is a modified Bessel function of the second kind. Using the asymptotic expansion of this function~\citep{Abramowitz1972}
\begin{equation}
\lim_{x\rightarrow \infty}K_{\alpha}(x)=\sqrt{\frac{\pi}{2x}}{\rm e}^{-x}\left(1+\sum_{n=1}^{\infty}\frac{\prod_{k=1}^n(4\alpha^2-(2k-1)^2)}{n!(8x)^n}\right),
\end{equation}
we obtain
\begin{align}
\mathcal{I}_1^{a>b} \approx{}& -\sqrt{\frac{\pi}{2}}\sum_{q=1}^{\infty}\left[\frac{1}{\sqrt{aq}}\left(-{\rm e}^{b-a}\right)^q+\frac{3}{8(aq)^{3/2}}\left(-{\rm e}^{b-a}\right)^q-\frac{15}{128(aq)^{5/2}}\left(-{\rm e}^{b-a}\right)^q+\frac{105}{1024(aq)^{7/2}}\left(-{\rm e}^{b-a}\right)^q+...\right]
\nonumber
\\
={}& -\sqrt{\frac{\pi}{2a}}\left[{\rm Li}_{1/2}\left(-{\rm e}^{b-a}\right)+\frac{3}{8a}{\rm Li}_{3/2}\left(-{\rm e}^{b-a}\right)-\frac{15}{128a^2}{\rm Li}_{5/2}\left(-{\rm e}^{b-a}\right)+\frac{105}{1024a^3}{\rm Li}_{7/2}\left(-{\rm e}^{b-a}\right)\right].
\end{align}
where ${\rm Li}_n(x)$ is the polylogarithm. Analogously for $\mathcal{I}_2$ we find
\begin{equation}
\mathcal{I}_2^{a>b}\approx \sqrt{\frac{\pi a}{2}}\left[{\rm Li}_{-1/2}\left(-{\rm e}^{b-a}\right)-\frac{1}{8a}{\rm Li}_{1/2}\left(-{\rm e}^{b-a}\right)+\frac{9}{128a^2}{\rm Li}_{3/2}\left(-{\rm e}^{b-a}\right)-\frac{75}{1024a^3}{\rm Li}_{5/2}\left(-{\rm e}^{b-a}\right)\right].
\end{equation}
We use these forms for $|b-a|<50$ if $\sqrt{a^2+b^2}>50$. For larger $a$, the Taylor expansion of the polylogarithm for ${\rm e}^{b-a}\ll 1$ is sufficiently accurate, and we write
\begin{align}
\mathcal{I}_1^{a\gg b}\simeq \sqrt{\frac{\pi}{2a}}{\rm e}^{b-a}\left(1+\frac{3}{8a}-\frac{15}{128a^2}+\frac{105}{1024a^3}\right),
\\
\mathcal{I}_2^{a\gg b}\simeq -\sqrt{\frac{\pi a}{2}}{\rm e}^{b-a}\left(1-\frac{1}{8a}+\frac{9}{128a^2}-\frac{75}{1024a^3}\right).
\end{align}

The analogous expansions/approximations for the other required partial derivatives are listed below:
\begin{align}
\left(\frac{\partial\Omega_{\rm e}}{\partial B}\right)_{n<n_{\text{max}}} = {}& \frac{e}{2\pi^2}\left(\Omega_B^{(-1)}(n')-\Omega_B^{(-1)}(1)\right) + \frac{e}{4\pi^2}\left(\Omega_B(0)+\Omega_B(n')+\Omega_B(1)\right) + \mathcal{B}_2\frac{e}{4\pi^2}\left(\Omega'_B(n')-\Omega'_B(1)\right) 
\nonumber
\\
{}& + \mathcal{B}_4\frac{e}{48\pi^2}\left(\Omega'''_B(n')-\Omega'''_B(1)\right) + ...,
\\
\Omega_B^{(-1)}(n) \equiv {}& -\frac{1}{8eB}\left[8eBn\mue E_{Fn}+m_{\rm e}^4{\rm arcoth}\left(\frac{\mue}{E_{Fn}}\right)-\left(m_{\rm e}^2+4eBn\right)^2\ln\left(\frac{\mue+E_{Fn}}{m_n}\right)\right],
\nonumber
\\
\Omega_{B}(n) \equiv {}& -\mue E_{Fn} + (m_{\rm e}^2+4eBn)\ln\left(\frac{\mue+E_{Fn}}{m_n}\right),
\nonumber
\\
\Omega'_{B}(n) \equiv {}& -2eB\left[\frac{eBn\mue}{m_n^2E_{Fn}} - 2\ln\left(\frac{\mue+E_{Fn}}{m_n}\right)\right],
\nonumber
\\
\Omega'''_{B}(n) \equiv {}& -\frac{2(eB)^3\mue}{m_n^6E_{Fn}^5}\left[20\mue^2m_n^2(m_{\rm e}^2+eBn)-3m_n^2(4m_{\rm e}^2+3eBn)-8\mue^4(m_{\rm e}^2+eBn)\right],
\nonumber
\\
\left(\frac{\partial\Omega_{\rm e}}{\partial B}\right)_{n\geq n_{\text{max}}} ={}& -\frac{e}{2\pi^2}\sum_{n=n_{\rm max}}^{n_{\rm max}+1}g_n\left[m_n^2\mathcal{G}_1(a,b)-(m_n^2+eBn)\mathcal{G}_2(a,b)\right],
\end{align}

\begin{align}
\left(\frac{\partial^2\Omega_{\rm e}}{\partial B^2}\right)_{n<n_{\text{max}}} = {}&
 \frac{e^2}{\pi^2}\left[n^{\prime 2}\ln\left(\frac{\mue+E_{Fn'}}{m_{n'}}\right)-\ln\left(\frac{\mu_{\rm e}E_{F1}}{m_1}\right)\right] + \frac{e^2}{2\pi^2}\left(\Omega_{BB}(n')+\Omega_{BB}(1)\right) + \mathcal{B}_2\frac{e^2}{2\pi^2}\left(\Omega'_{BB}(n')+\Omega'_{BB}(1)\right)
\nonumber
\\
{}& + \mathcal{B}_4\frac{e^2}{24\pi^2}\left(\Omega'''_{BB}(n')+\Omega'''_{BB}(1)\right) + ...,
\\
\Omega_{BB}(n) \equiv {}& -\frac{eBn^2\mue}{m^2_nE_{Fn}} + 2n\ln\left(\frac{\mu_{\rm e}+E_{Fn}}{m_n}\right),
\nonumber
\\
\Omega'_{BB}(n) \equiv {}& \frac{eBn\mue\left[eBn(5m_n^2+2\mue^2)-4m_n^4E_{Fn}^2\right]}{m_n^4E^3_{Fn}} + 2\ln\left(\frac{\mue+E_{Fn}}{m_n}\right),
\nonumber
\\
\Omega'''_{BB}(n) \equiv {}& \frac{(eB)^2\mue}{m_n^8E^7_{Fn}}\Big[12m_n^4(2\mue^2-3m_n^2)E_{Fn}^4 - 8eBnm_n^2E_{Fn}^2(8\mue^4-20m_n^2\mue^2+15m_n^4)
\nonumber
\\
{}& \qquad\qquad\quad +3(eBn)^2(16\mue^6-56\mue^4m_n^2+70\mue^2m_n^4-35m_n^6)\Big],
\nonumber
\\
\left(\frac{\partial^2\Omega_{\rm e}}{\partial B^2}\right)_{n\geq n_{\rm max}} \approx {}& \frac{e^2}{2\pi^2}\sum_{n=n_{\rm max}}^{n_{\rm max}+1}g_nn\left(2\mathcal{G}_2(a,b)+\frac{eBn}{m_n^2}\mathcal{H}_2(a,b)\right),
\end{align}
where
\begin{subequations}
\begin{align}
\mathcal{G}_1(a,b)\equiv{}& \int_{1}^{\infty}{\rm d}x\frac{x^2}{\sqrt{x^2-1}}\frac{1}{{\rm e}^{ax-b}+1},
\\
\mathcal{G}_2(a,b)\equiv{}& \int_{1}^{\infty}{\rm d}x\frac{1}{\sqrt{x^2-1}}\frac{1}{{\rm e}^{ax-b}+1},
\\
\mathcal{H}_2(a,b)\equiv{}& -a\int_{1}^{\infty}{\rm d}x\frac{x}{\sqrt{x^2-1}}\frac{{\rm e}^{ax-b}}{\left({\rm e}^{ax-b}+1\right)^2},
\end{align}
\end{subequations}
which have asymptotic forms
\begin{subequations}
\begin{align}
\mathcal{G}^{a>b}_1\approx{}& -\sqrt{\frac{\pi}{2a}}\left[{\rm Li}_{1/2}\left(-{\rm e}^{b-a}\right)+\frac{7}{8a}{\rm Li}_{3/2}\left(-{\rm e}^{b-a}\right)+\frac{57}{128a^2}{\rm Li}_{5/2}\left(-{\rm e}^{b-a}\right)-\frac{195}{1024a^3}{\rm Li}_{7/2}\left(-{\rm e}^{b-a}\right)\right],
\\
\mathcal{G}^{a\gg b}_1\approx{}& \sqrt{\frac{\pi}{2a}}{\rm e}^{b-a}\left[1+\frac{7}{8a}+\frac{57}{128a^2}-\frac{195}{1024a^3}\right],
\\
\mathcal{G}^{a>b}_2\approx{}& -\sqrt{\frac{\pi}{2a}}\left[{\rm Li}_{1/2}\left(-{\rm e}^{b-a}\right)-\frac{1}{8a}{\rm Li}_{3/2}\left(-{\rm e}^{b-a}\right)+\frac{9}{128a^2}{\rm Li}_{5/2}\left(-{\rm e}^{b-a}\right)-\frac{75}{1024a^3}{\rm Li}_{7/2}\left(-{\rm e}^{b-a}\right)\right],
\\
\mathcal{G}^{a\gg b}_2\approx{}& \sqrt{\frac{\pi}{2a}}{\rm e}^{b-a}\left[1-\frac{1}{8a}+\frac{9}{128a^2}-\frac{75}{1024a^3}\right],
\\
\mathcal{H}^{a>b}_2\approx{}& \sqrt{\frac{\pi a}{2}}\left[{\rm Li}_{-1/2}\left(-{\rm e}^{b-a}\right)+\frac{3}{8a}{\rm Li}_{1/2}\left(-{\rm e}^{b-a}\right)-\frac{15}{128a^2}{\rm Li}_{3/2}\left(-{\rm e}^{b-a}\right)+\frac{105}{1024a^3}{\rm Li}_{5/2}\left(-{\rm e}^{b-a}\right)\right], 
\\
\mathcal{H}^{a\gg b}_2\approx{}& -\sqrt{\frac{\pi a}{2}}{\rm e}^{b-a}\left[1+\frac{3}{8a}-\frac{15}{128a^2}+\frac{105}{1024a^3}\right].
\end{align}
\end{subequations}
For $b-a>50$ we use the forms
\begin{align}
\left(\frac{\partial\Omega_{\rm e}}{\partial B}\right)^{b\gg a}_{n\geq n_{\text{max}}} ={}& -\frac{e}{2\pi^2}\sum_{n=n_{\text{max}}}^{n_{\text{max}}+1}g_n\Bigg[\frac{1}{2}\mue\sqrt{\mue^2-m_n^2}-\frac{1}{2}(m_n^2+2eBn)\ln\left(\frac{\mue+\sqrt{\mue^2-m_n^2}}{m_n}\right)
+\frac{\pi^2}{6}\frac{\mue(\mue^2-m^2_{\rm e}-eBn)}{(\mue^2-m_n^2)^{3/2}}T^2
\nonumber
\\
{}& \qquad\qquad\qquad\quad +\frac{7\pi^4}{120}\frac{\mue(\mue^2(m_{\rm e}^2+4eBn)+2(eBn)^2)-m^2_{\rm e}(m^2_{\rm e}+eBn)}{(\mue^2-m_n^2)^{7/2}}T^4\Bigg],
\\
\left(\frac{\partial^2\Omega_{\rm e}}{\partial B^2}\right)^{b\gg a}_{n\geq n_{\text{max}}} ={}& \frac{e^2}{2\pi^2}\sum_{n=n_{\text{max}}}^{n_{\text{max}}+1}g_nn\Bigg[-\frac{eBn\mue}{m_n^2\sqrt{\mue^2-m_n^2}}+2\ln\left(\frac{\mue+\sqrt{\mue^2-m_n^2}}{m_n}\right)
-\frac{\pi^2}{6}\frac{\mue(2\mue^2-2m^2_n+3eBn)}{(\mue^2-m_n^2)^{5/2}}T^2
\nonumber
\\
{}& \qquad\qquad\qquad\quad -\frac{7\pi^4}{120}\frac{\mue\left(4\mue^4-6m_n^4+2\mue^2m_n^2+(20\mue^2+15m_n^2)eBn\right)}{(\mue^2-m_n^2)^{9/2}}T^4\Bigg].
\end{align}

\subsection{High-temperature Approximations}
\label{app:HighTApprox}

At high temperatures $T> eB/(2\pi^2\mu_{\rm e})$, the approximations in the previous Appendix Section break down, as additional Landau levels above $n=\nmax+1$ must be included to properly account for nonzero temperatures. So in this region of parameter space we require a new approximate form for the required partial derivatives. 

As shown in~\citet{Elmfors1993}, the grand potential density $\Omega_{\rm e}$ (equal to the negative of the Lagrangian density $\Lagr$) can be split into three parts: a $B$-independent part $\Omega_{\rm e,0}$ that is the usual grand potential density for a nonmagnetized relativistic Fermi gas, and two $B$-dependent parts labeled $\Omega_{\rm e,\text{reg}}$ for ``regular'' and $\Omega_{\rm e,\text{osc}}$ for ``oscillatory''. The latter two are given by
\begin{align}
\Omega_{\rm e,\text{reg}}={}& -\frac{(eB)^{3/2}}{4\pi^{5/2}}\int^{\infty}_0\frac{{\rm d}y}{y^{5/2}}\left[y\text{coth}y-1\right]\int_{m_{\rm e}}^{\infty}{\rm d}Ef_F(E,\mue,T){\rm e}^{-y\frac{E^2-m^2_{\rm e}}{eB}} 
\xrightarrow{T=0}
-\frac{(eB)^{3/2}}{4\pi^{5/2}}\int^{\infty}_0\frac{{\rm d}y}{y^{5/2}}\left[y\text{coth}y-1\right]\int_{m_{\rm e}}^{\mu_{\rm e}}{\rm d}E{\rm e}^{-y\frac{E^2-m^2_{\rm e}}{eB}},
\\
\Omega_{\rm e,\text{osc}}={}& \frac{(eB)^{3/2}}{2\pi^3}\sum^{\infty}_{p=1}\frac{1}{p^{3/2}}\int^{\infty}_{m_{\rm e}}{\rm d}Ef_F(E,\mue,T)\sin\left[\frac{\pi}{4}-\frac{\pi p}{eB}(E^2-m^2_{\rm e})\right]
\xrightarrow{T=0}
 \frac{(eB)^{3/2}}{2\pi^3}\sum^{\infty}_{p=1}\frac{1}{p^{3/2}}\int^{\mu_{\rm e}}_{m_{\rm e}}{\rm d}E\sin\left[\frac{\pi}{4}-\frac{\pi p}{eB}(E^2-m^2_{\rm e})\right],
\label{eq:OmegaOsc}
\end{align}
for $f_F(E,\mue,T)=(\exp([E-\mue]/T)+1)^{-1}$. Eq.~(\ref{eq:OmegaOsc}) for $T=0$ can be approximated as
\begin{align}
\Omega_{\rm e,\text{osc}}\approx{}& \frac{(eB)^2}{4\pi^3}\sum_{p=1}^{\infty}\frac{1}{p^2}\Bigg[ \sin\left(\frac{\pi p m_{\rm e}^2}{eB}\right) + \cos\left(\frac{\pi p m_{\rm e}^2}{eB}\right)\left\{S\left(\sqrt{\frac{2pm_{\rm e}^2}{eB}}\right)-C\left(\sqrt{\frac{2pm_{\rm e}^2}{eB}}\right)\right\} - \sin\left(\frac{\pi p m_{\rm e}^2}{eB}\right)\left\{S\left(\sqrt{\frac{2pm_{\rm e}^2}{eB}}\right)+C\left(\sqrt{\frac{2pm_{\rm e}^2}{eB}}\right)\right\} \Bigg]
\nonumber
\\
{}&+\frac{(eB)^{5/2}}{4\pi^4\mue}\sum_{p=1}^{\infty}\frac{1}{p^{5/2}}\cos\left(\frac{\pi}{4}-\pi p\frac{p_F^2}{eB}\right),
\label{eq:OmegaeOsc}
\end{align}
where $C(x)$ and $S(x)$ are the Fresnel integrals defined as in~\citet{Abramowitz1972}, the asymptotic expansions of which were used assuming $p_F^2+m_{\rm e}^2=\mue^2\gg eB$ to obtain the last contribution to this equation. To a very good approximation, all but this final contribution can be dropped from Eq.~(\ref{eq:OmegaeOsc}).

We can include the temperature-dependence in this grand potential by: (1) using the Sommerfeld approximation for $\Omega_{\rm e,0}$ and (2) by applying a temperature-dependent correction factor to each term in $\Omega_{\rm e,\text{osc}}$ which avoids having to perform computationally expensive Fermi--Dirac integrals. Following~\citet{Shoenberg1984} and generalizing to the relativistic case by replacing the electron mass with its effective mass $\mue$, this correction factor is
\begin{equation}
R_{T,p} = \frac{2\pi^2p\mue T}{eB}\text{csch}\left(\frac{2\pi^2p\mue T}{eB}\right),
\end{equation}
inserted into the $p^{\text{th}}$ term in the sum in $\Omega_{\rm e,\text{osc}}$. For high temperatures and many occupied Landau levels $\mu_{\rm e}\gg \sqrt{eB}$, this factor suppresses the higher-order terms in the sum, to an extent that including the $p=1$ term alone can be an excellent approximation. Since $\text{csch}[2\pi^2p\mue T/(eB)]\approx\exp[-2\pi^2p\mue T/(eB)]$ for high temperatures, this factor is the origin of $T=eB/(2\pi^2\mue)$ as the ``cutoff'' between high- and low-temperature regimes. We can cut off the sum over $p$ at $p=5$ and obtain very accurate results in the parameter range of interest.

The high-temperature approximations for the necessary partial derivatives are calculated by taking partial derivatives of $\Omega_{\rm e,0}$, $\Omega_{\rm e,\text{reg}}$ and $\Omega_{\rm e,\text{osc}}$. Nonzero temperature is accounted for by including the nonzero temperature correction factors in $\Omega_{\rm e,\text{osc}}$, through Sommerfeld expansion terms for $\Omega_{\rm e,0}$ where they are nonzero, and Sommerfeld expansion terms for partial derivatives of $\Omega_{\rm e,\text{reg}}$ with respect to $T$. The resulting equations are
\begin{subequations}
\begin{align}
\frac{\partial\Omega_{\rm e}}{\partial B} \approx {}& \frac{e\sqrt{eB}}{4\pi^3\mue}\sum_{p=1}^{5}\frac{R_{T,p}}{p^{3/2}}\left[-p_F^2\sin\left(\frac{\pi}{4}-\pi p\frac{p_F^2}{eB}\right)+\frac{5 eB}{2\pi p}\cos\left(\frac{\pi}{4}-\pi p\frac{p_F^2}{eB}\right)\right]
\nonumber
\\
{}& -\frac{e\sqrt{eB}}{2\pi^{5/2}}\left[\mathcal{J}_0(\mue,eB)-\frac{m_{\rm e}^2}{2eB}\mathcal{J}_1(\mue,eB)-\frac{1}{4}\left(\mue\mathcal{K}_0\left(\frac{p_F^2}{eB}\right)-m_{\rm e}\mathcal{K}_0(0)\right)\right]
\\
\frac{\partial^2\Omega_{\rm e}}{\partial B^2} \approx {}& -\frac{e^2p_F^2}{4\pi^2\mue\sqrt{eB}}\sum_{p=1}^{5}\frac{R_{T,p}}{\sqrt{p}}\left[\frac{p_F^2}{eB}\cos\left(\frac{\pi}{4}-\pi p\frac{p_F^2}{eB}\right)+\frac{3}{\pi p}\sin\left(\frac{\pi}{4}-\pi p\frac{p_F^2}{eB}\right)\right]
\nonumber
\\
{}& -\frac{e^2}{2\pi^{5/2}\sqrt{eB}}\left[\mathcal{J}_0(\mue,eB) - \frac{m_{\rm e}^2}{eB}\mathcal{J}_1(\mue,eB)  - \frac{5}{8}\left\{\mue\mathcal{K}_0\left(\frac{p_F^2}{eB}\right)-m_{\rm e}\mathcal{K}_0(0)\right\} \right],
\\
\frac{\partial^2\Omega_{\rm e}}{\partial B\partial\mue} \approx {}& \frac{e}{2\pi^2\sqrt{eB}}\sum_{p=1}^{5}\frac{R_{T,p}}{\sqrt{p}}\left[p_F^2\cos\left(\frac{\pi}{4}-\pi p\frac{p_F^2}{eB}\right)+\frac{3eB}{2\pi p}\sin\left(\frac{\pi}{4}-\pi p\frac{p_F^2}{eB}\right)\right] - \frac{1}{4\pi^{5/2}\sqrt{eB}}\left[p_F^2\mathcal{K}_1\left(\frac{p_F^2}{eB}\right) + \frac{3eB}{2}\mathcal{K}_0\left(\frac{p_F^2}{eB}\right)\right],
\end{align}
\end{subequations}
where we retain only the leading order or a few subleading order terms in the ratio $p_F^2/eB\gg 1$ and define the auxiliary functions
\begin{align}
\mathcal{J}_p(\mue,eB) ={}& \int_0^{\infty}\frac{{\rm d}y}{y^{5/2-p}}\left[y\text{coth}y-1\right]\int_{m_{\rm e}}^{\mue}{\rm d}E{\rm e}^{-y\frac{E^2-m_{\rm e}^2}{eB}},
\\
\mathcal{K}_p(x) ={}& \int_0^{\infty}\frac{{\rm d}y}{y^{5/2-p}}\left[y\text{coth}y-1\right]{\rm e}^{-yx}.
\end{align}

The approximations for each of the three required partial derivatives were compared to their exact forms, computed by calculating the full Fermi-Dirac integrals. Over a parameter space range $10^{14}\text{ G}<B<5\times10^{16}$ G, $10\text{ MeV}<\mue<80$ MeV and $5\times10^{7}\text{ K}<T<5\times10^9$ K, the mean relative error between the approximations and exact values is $\lesssim $1\% for all three functions $M$, $\chi_{\mu}$ and $\mathcal{M}_{\mu}$. The maximum relative error between approximation and exact values is 4\% (for $M$ and $\mathcal{M}_{\mu}$) and 8\% ($\chi_{\mu}$).

\bibliographystyle{aasjournal}
\bibliography{library,textbooks,librarySpecial}

\end{document}